\documentclass[useAMS,usenatbib]{mnras/mnras}

\usepackage{amssymb}
\usepackage{graphicx}
\usepackage[usenames,dvipsnames]{color}
\usepackage{verbatim}

\usepackage{times}
\usepackage{mathptmx}

\newcommand{\um}[1]{\ \mathrm{#1}}
\newcommand{\mic}[1]{\ \umu\mathrm{#1}}

\newcommand{\xx}{\textcolor{red}{xx}}

\newcommand{\hii}{H \textsc{ii} region}
\newcommand{\ObjTot}{99} 
\newcommand{\SrcDis}{70} 
\newcommand{\SrcDif}{29} 

\title[Extended Galactic sources in SCORPIO]{Study of the Galactic radio sources in the SCORPIO survey resolved by ATCA at 2.1 GHz}
\author[A. Ingallinera et al.]{{\small A. Ingallinera$^1$\thanks{E-mail:ingallinera@oact.inaf.it}, G. Umana$^1$, C. Trigilio$^1$, R. Norris$^{2,3}$, T.M.O. Franzen$^{3,4}$, F. Cavallaro$^1$, P. Leto$^1$, C. Buemi$^1$, F. Schillir\`o$^1$, F. Bufano$^1$, S. Riggi$^1$, S. Loru$^1$, C. Agliozzo$^5$}\\
$^1$INAF-Osservatorio Astrofisico di Catania, Via Santa Sofia 78, 95123 Catania, Italy\\
$^2$Western Sydney University, Locked Bag 1797, Penrith South, NSW 1797, Australia\\
$^3$CSIRO Astronomy and Space Science, PO Box 1130, Bentley, WA 6102, Australia\\
$^4$ASTRON, Netherlands Institute for Radio Astronomy, Oude Hoogeveensedijk 4, 7991 PD, Dwingeloo, The Netherlands\\
$^5$European Southern Observatory, Karl-Schwarzschild-Strasse 2, Garching bei M\"unchen, 85748, Germany
}

\begin{document}

\date{Accepted 2019 October 18. Received 2019 October 11; in original form 2019 August 09}

\pagerange{\pageref{firstpage}--\pageref{lastpage}} \pubyear{2002}

\maketitle

\label{firstpage}

\begin{abstract}
We present a catalogue of a large sample of extended radio sources in the SCORPIO field, observed and resolved by the Australia Telescope Compact Array. SCORPIO, a pathfinder project for addressing the early operations of the Australia SKA Pathfinder, is a survey of $\sim\!5$ square degrees between 1.4 and $3.1\um{GHz}$, centered at $l=343.5^\circ$, $b=0.75^\circ$ and with an angular resolution of about $10\um{arcsec}$. It is aimed at understanding the scientific and technical challenges to be faced by future Galactic surveys. With a mean sensitivity around $100\mic{Jy\ beam}^{-1}$ and the possibility to recover angular scales at least up to $4\um{arcmin}$, we extracted 99 extended sources, 35 of them detected for the first time. Among the 64 known sources 55 had at least a tentative classification in literature. Studying the radio morphology and comparing the radio emission with infrared we propose as candidates 6 new \hii s, 2 new planetary nebulae, 2 new luminous blue variable or Wolf--Rayet stars and 3 new supernova remnants. This study provides an overview of the potentiality of future radio surveys in terms of Galactic source extraction and characterization and a discussion on the difficulty to reduce and analyze interferometric data on the Galactic plane.

\end{abstract}


\section{Introduction}
\label{sec:intro}
The advent of next-generation radio instruments, like the Square Kilometre Array (SKA), is going to revolutionize our knowledge about Galactic objects. High resolution, unprecedented sensitivity and the possibility of almost-all-sky surveys are going to supply a huge quantity of top-level data, shedding light on poorly studied or understood phenomena. Beside the ubiquitous Galactic diffuse emission (GDE; e.g. \citealt{Zheng2017}), thousands of known discrete sources constitute the radio Galactic zoo, most notably H \textsc{ii} regions, supernova remnants (SNRs), planetary nebulae (PNe), luminous blue variables (LBVs) and radio stars. Although these objects have been the subject of multiwavelength studies for several decades, different open issues remain. For example many details of the stellar evolution are still under debate, especially the last phases of massive stars \citep{Umana2011}, and it is expected that most of the Galactic SNRs and PNe have not been observed yet \citep{Dubner2015,Sabin2014}. Finally the significant sensitivity improvement will make other categories of Galactic objects, like quiescent main-sequence stars, detectable.

Waiting for SKA to become operational, its precursors have already been built and tested. The Australian SKA Pathfinder (ASKAP), for example, entered its early-science phase in late 2017. One of the large surveys already approved for ASKAP is the Evolutionary Map of the Universe (EMU; \citealt{Norris2011}), scheduled to start in mid 2019. The main goal of EMU is to make a deep ($\mathrm{rms}\!\sim\!10\mic{Jy\ beam}^{-1}$) continuum survey of the entire southern sky at $1.3\um{GHz}$, extending up to a declination of $+30^\circ$, with a resolution of $10\um{arcsec}$. EMU will observe also $\sim\!75$ percent of the Galactic plane, creating the most sensitive wide-field atlas of Galactic continuum emission yet made in the southern hemisphere. Since EMU is the first survey to observe a so large area of the sky with this sensitivity, to achieve its ambitious goal it will have to face new challenges on how to reduce and analyze the huge quantity of data. A current challenge is to predict what we should expect to detect. For this reason, in the wide context of the preparation for EMU, we started the `Stellar Continuum Originating from Radio Physics In Ourgalaxy' (SCORPIO) project, a blind survey of an approximately $2\times2$-deg$^2$ area of sky centred at Galactic coordinates $l=343.5^\circ$, $b=0.75^\circ$ (\citealt{Umana2015}; hereafter `Paper I'), originally observed with the Australia Telescope Compact Array (ATCA). The major scientific goals of SCORPIO are the production of catalogues of different populations of Galactic radio point sources and the study of circumstellar envelopes (related to young or evolved massive stars, PNe and SNRs) which is extremely important for understanding the Galaxy evolution (e.g. interstellar medium chemical enrichment, star formation triggering, etc.). Besides these scientific outputs, SCORPIO will be used as a technical test-bed for the EMU survey, helping to shape the strategy for its Galactic plane sections. SCORPIO results are, in fact, meant to guide EMU design in: identifying issues arising from the complex structures; identifying issues arising from the variable sources; testing which is the most appropriate method of finding and extracting sources embedded in the diffuse emission expected at low Galactic latitude; quantifying how effectively we can identify counterparts to the extracted sources and disentangle different radio source populations.

Imaging and analyzing extended sources with interferometers is particularly problematic. Firstly the most wide-spread deconvolution algorithms are designed and optimized for point sources. Secondly, source dimension can exceed the interferometer largest angular scale (LAS), resulting in imaging artifacts and flux-loss. On the other hand, one of the highest impact products of future deep Galactic survey with arcsecond or sub-arcsecond resolution will be the observation of thousands of Galactic extended sources. The sensitivity improvement will allow the detection of those extended sources characterized by a low brightness.

This work presents a catalogue of all the extended (in the sense defined in Section \ref{sec:detcla}) sources in the SCORPIO field. Further goals are: to find the best method to reduce and analyze large radio dataset toward the Galactic plane; to assess what kind of and how many Galactic objects we expect to find in future surveys; to predict which physical information we can derive from this kind of radio observations. In Section \ref{sec:obsred} we describe the observations and the data reduction. In Section \ref{sec:detcla} we report the source extraction procedure and the general classification scheme that we used to compile the whole catalogue (see Appendix A). In Sections \ref{sec:hii}, \ref{sec:pn} and \ref{sec:mass} we discuss individually each class of Galactic source. A final discussion and the conclusions are reported in Sections \ref{sec:dis} and \ref{sec:con}.

\section{Observations and data reduction}
\label{sec:obsred}

\subsection{ATCA observations}
\label{sec:obs}
The SCORPIO field was observed with ATCA for a total of $\sim\!320$ hours split in different blocks from 2011 to 2016. The observations were made using the Compact Array Broad-band Backend \citep{Wilson2011}, with an effective observing band of $2\um{GHz}$, divided into 2048 1-MHz channels, between 1.1 and $3.1\um{GHz}$. All four Stokes parameters were measured and polarization studies will be the subject of future papers. Because of strong RFI corruption the useful band was reduced to the range $1.4-3.1\um{GHz}$. The first observations on April 2011 and June 2012, in 6A and 6B configuration (`the extended configurations'), covered 1/4 of the field, the so-called `pilot region', and are discussed in detail in Paper I. On June and August 2012 observations in extended configuration (respectively 6B and 6A) covered the entire field. The field was observed extending the hexagonal pointing grid of the pilot region, with a total of 95 new pointings. Considering the 38 pointings of the pilot region, the entire SCORPIO field was then covered with 133 pointings, with a constant pointing spacing of $8.8\um{arcmin}$. The primary beam ranges from $33\um{arcmin}$ at $1.4\um{GHz}$ to $15\um{arcmin}$ at $3.1\um{GHz}$. A total of 192 hours were spent in extended configurations. Taking into account overheads due to calibration and antenna moving, each pointing was observed for about 1.2 hours. The main reason behind the use of the extended configurations was to achieve the highest possible resolution with ATCA. However the coverage of the $uv$ plane, especially toward its centre, was very poor. As a result, imaging artifacts heavily affected the final maps, with extended sources totally filtered out and a ten-fold increase of the background noise toward the Galactic plane. Considering an observing frequency band between 1.4 and $3.1\um{GHz}$ the maximum theoretical value for the LAS ranges from $\sim\!3.4$ to $\sim\!1.6\um{arcmin}$. Taking into account the imperfect $uv$ coverage, these values are to be viewed as upper limits. Therefore, sources much more extended than about $1\um{arcmin}$ could not be properly imaged with extended configurations across the whole observing band.

\begin{figure*}
	\includegraphics[height=6cm]{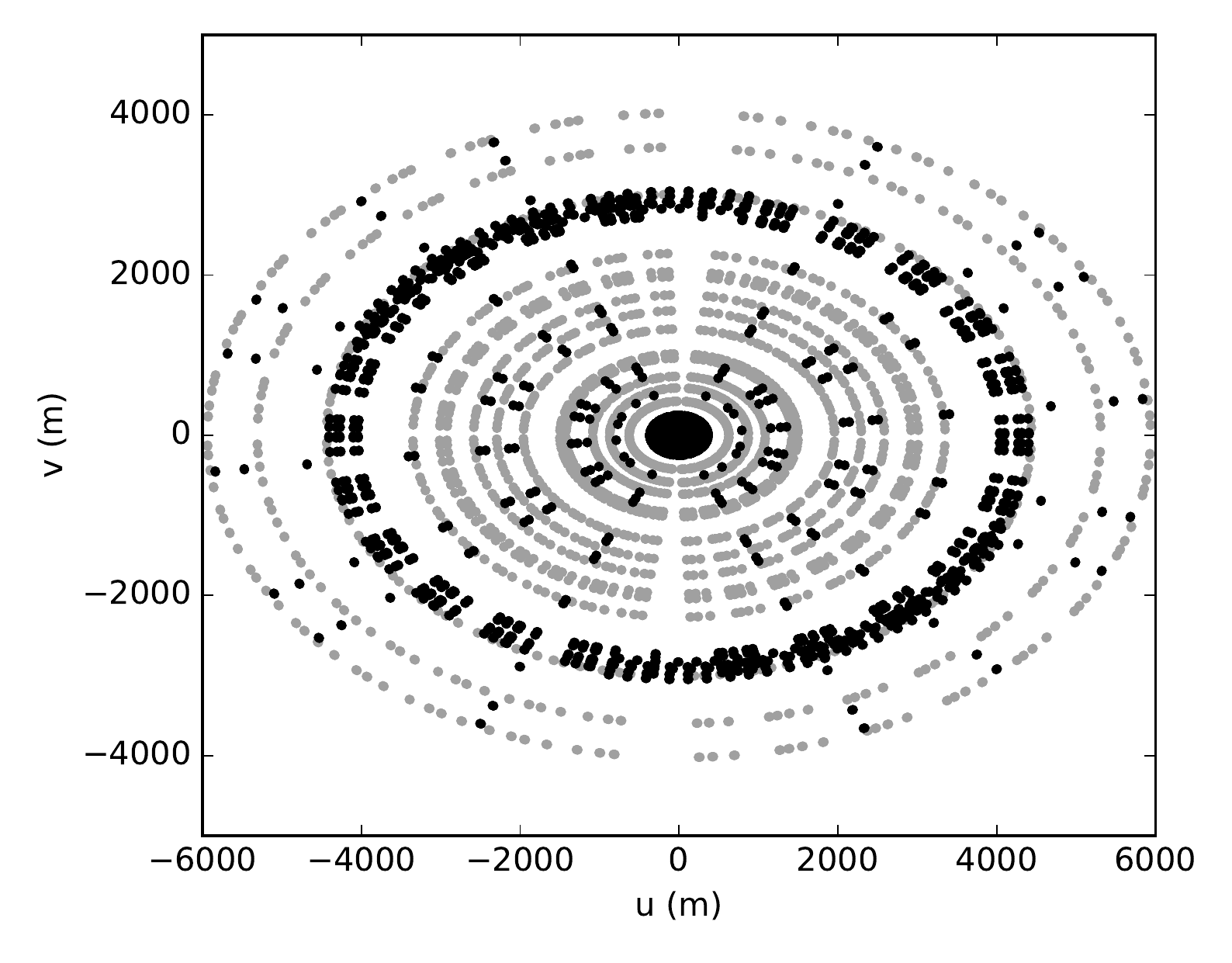}\hspace{1cm}
    \includegraphics[height=6cm]{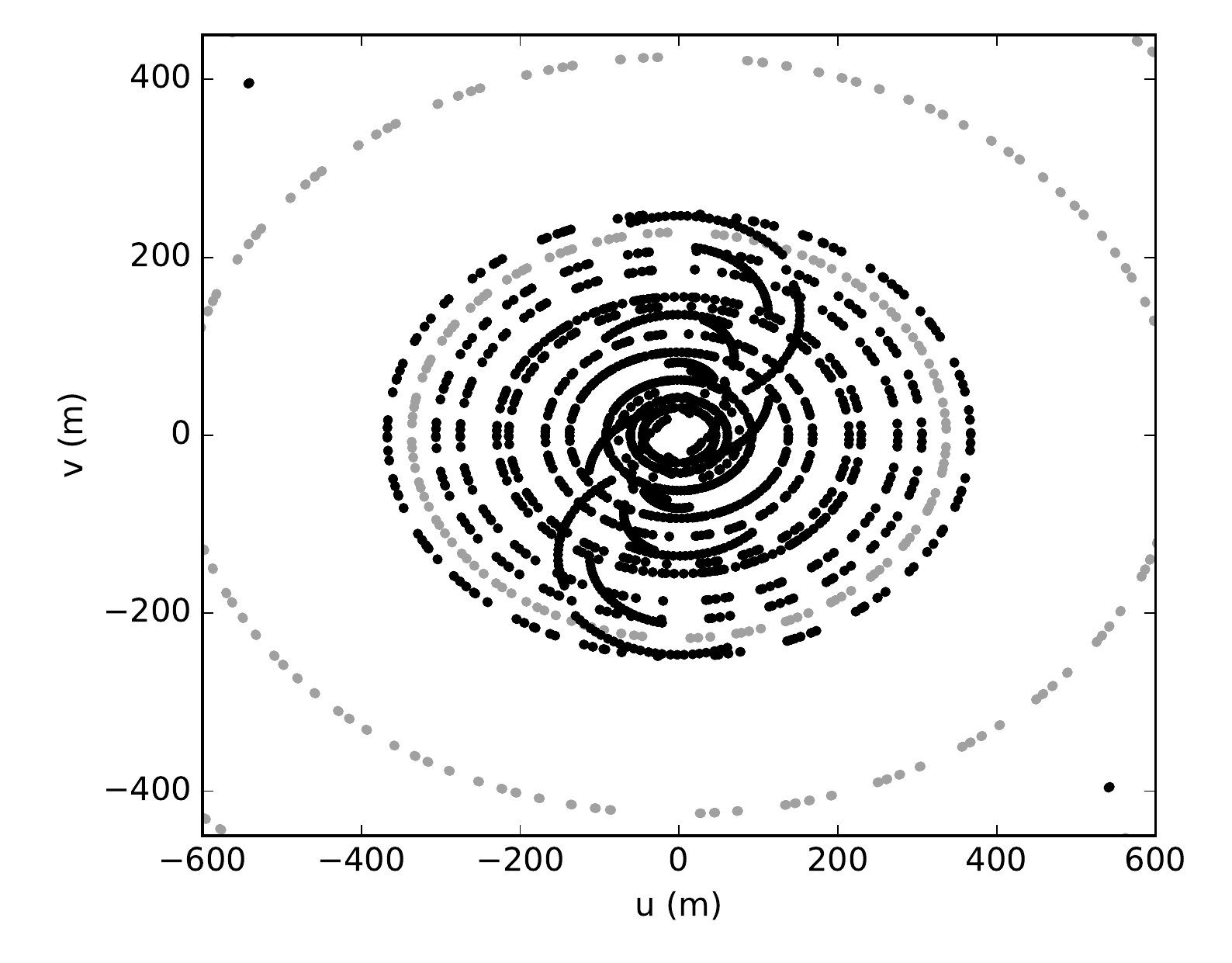}
	\caption{Coverage of the $uv$-plane for a single pointing (all the pointings have a very similar coverage). In the left panel the complete $uv$ plane, with gray dots referring to extended-configuration observations and black dots to compact-configuration. In the right panel a zoom of the inner part of the same plot, highlighting the significant improvement of the coverage in the central part of the $uv$ plane.}
	\label{fig:uvcov}
\end{figure*}

The SCORPIO $uv$ coverage was subsequently improved in February 2014 and January/February 2016 using configurations EW367, EW352 and H214 (`the compact configurations'). These observations covered the entire field and were aimed to test what kind of improvements short baselines would have introduced. Combining these data with the previous observations in extended configuration, the signal-to-noise ratio actually increased in the regions where extended or diffuse emission is present, thanks to the reduction of the artifacts deriving from imperfect cleaning of the resolved-out sources. After this concatenation the LAS at $2.1\um{GHz}$ is prudently of order of $4\um{arcmin}$ (theoretical calculation returns a value of about $10\um{arcmin}$, but the shortest-baseline data were the most affected by flagging). The $uv$-plane coverage of a single pointing, comprising both configurations, is reported in Figure \ref{fig:uvcov}.

\subsection{Data reduction}
\label{sec:red}
The data reduction process was mainly conducted with \textsc{miriad} \citep{Sault1995}. The procedure was slightly modified with respect to that presented in Paper I only in the imaging part, to take into account the presence of extended sources and diffuse emission. Regarding the datasets derived from extended-configuration observations, the editing and calibration procedure was the same as discussed in Paper I and will be reported in detail in (Trigilio et al. \textit{in prep.}).

Regarding compact-configuration observations, as a first step the datasets were flagged for bad data using the tasks \textsc{pgflag} and \textsc{uvflag}. The bandpass calibration was performed with the task \textsc{mfcal} using as calibrator the standard source 1934-638. The gain calibration was subsequently performed with \textsc{gpcal} and using as calibrator the source 1714-397. The datasets were finally scaled for flux density using again the standard source 1934-638, with a flux density of $12.31\um{Jy}$ at $2.1\um{GHz}$ \citep{Reynolds1994}.

As stated in Section \ref{sec:obs}, all the pointings performed in compact configuration had the same phase centre as those in extended configuration. Therefore, once extended- and compact-configuration data were calibrated, they were combined in the $uv$-plane using the task \textsc{uvaver}. As a result we obtained 133 visibility datasets, one for each pointing, which contained data deriving from both configuration observations.

For the imaging process we decided to test two different approaches. The first approach was similar to that presented in Paper I. Each pointing was imaged separately (using tasks \textsc{invert}, \textsc{mfclean} and \textsc{restore}) and then a complete map was obtained mosaicing all the pointings together (task \textsc{linmos}). The second approach was to use the maximum entropy method (task \textsc{mosmem}).

\begin{figure*}
	\includegraphics[height=6cm]{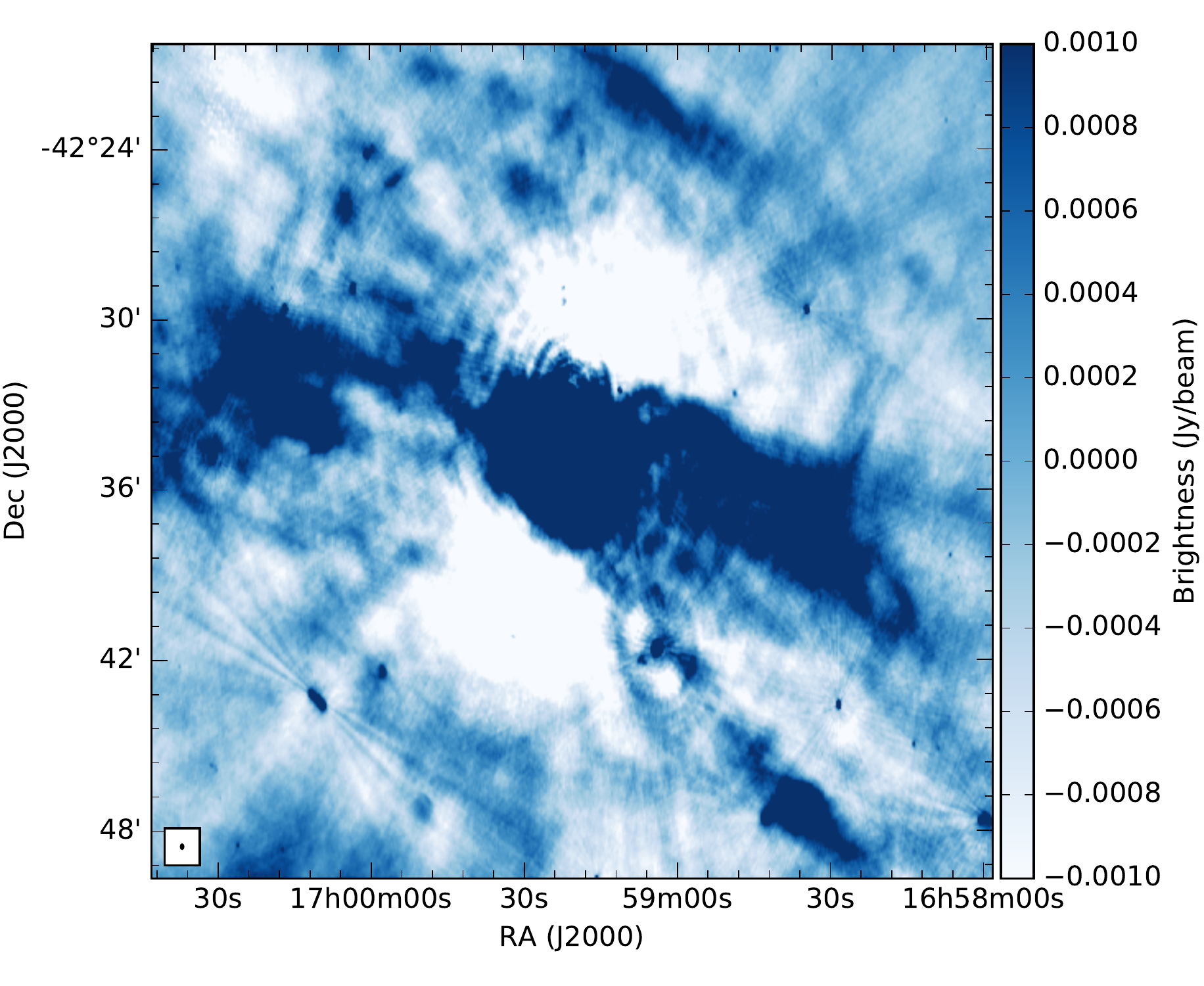}\hspace{1cm}
    \includegraphics[height=6cm]{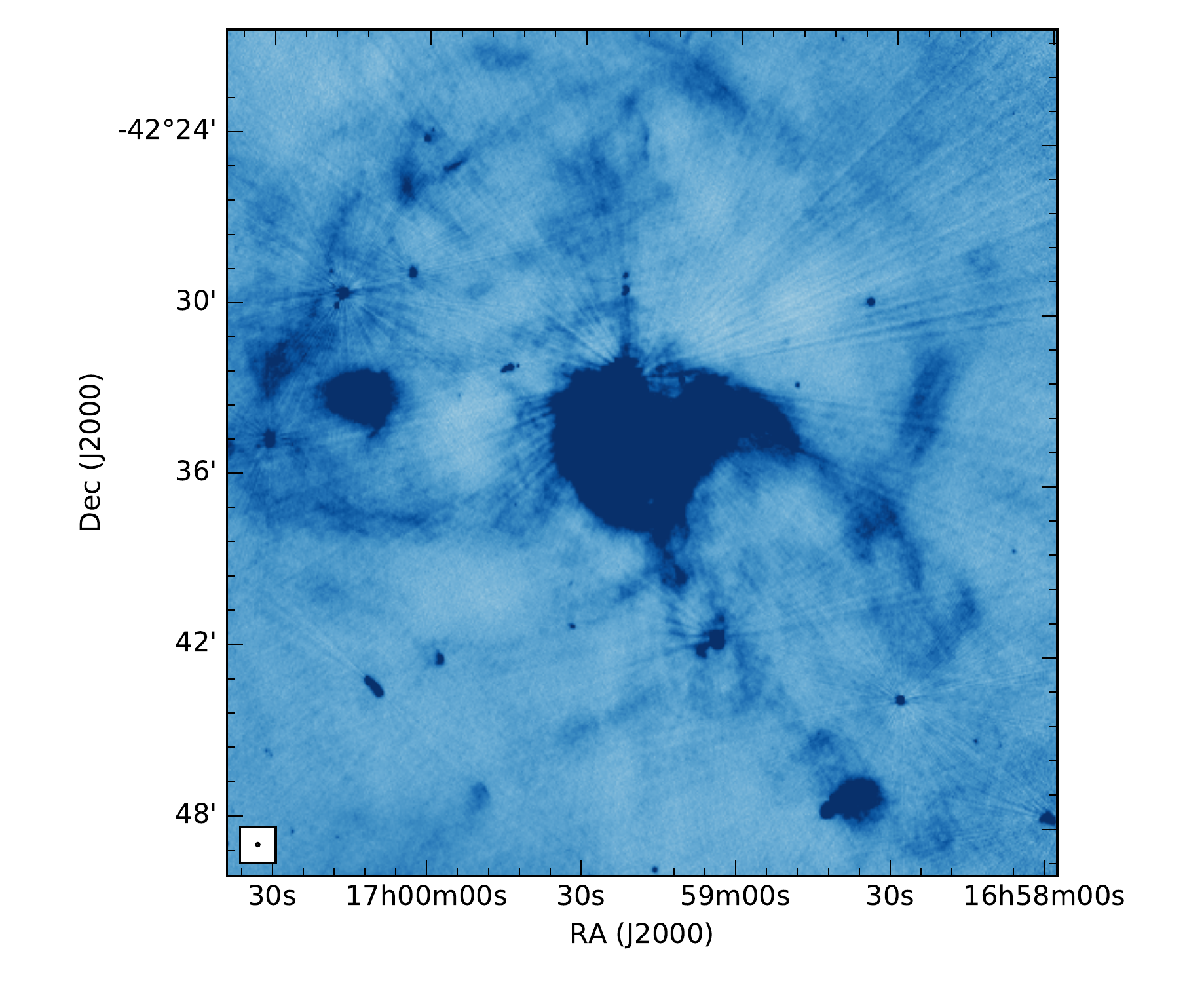}
	\caption{A comparison of the same region of the SCORPIO field extracted from the `\textsc{linmos}' map (left) and from the `\textsc{mosmem}' map (right). The color scale is the same for both images. The `\textsc{mosmem}' map shows a more regular background, without the prominent artifacts of the `\textsc{linmos}' map. Also the low-brightness diffuse emission seems to be better recovered in the `\textsc{mosmem}' map. The peak brightness for both maps is around $160\um{mJy\ beam}^{-1}$, but the standard deviation measured close to the bright extended source in the center is $0.87\um{mJy\ beam}^{-1}$ for the `\textsc{linmos}' map and $0.08\um{mJy\ beam}^{-1}$ for the `\textsc{mosmem}' map. Radial artifacts around bright point sources are likely due to residual unflagged bad data.}
	\label{fig:linmem}
\end{figure*}

At the end we created two maps, we call them the `\textsc{linmos}' map (synthesized beam $9\times5\um{arcsec}^2$) and the `\textsc{mosmem}' map (synthesized beam $6.5\times6.5\um{arcsec}^2$). The two maps show significant differences on how extended sources are imaged, with the `\textsc{mosmem}' map giving the best aesthetic result. The `\textsc{mosmem}' map gives also an almost uniform background noise, with an rms around $100\mic{Jy\ beam}^{-1}$ almost everywhere. On the other hand the `\textsc{linmos}' map, worse in reducing imaging artifacts, is characterized by a variable background noise, with an rms around $30\mic{Jy\ beam}^{-1}$ far from extended sources but with an increase up to $400\mic{Jy\ beam}^{-1}$ close to the Galactic plane. So \textsc{linmos} seems to be best suitable for fields where no extended sources are present, while \textsc{mosmem} in the opposite case. In a field like SCORPIO, where regions crowded and devoid of extended sources co-exist, there is not an imaging procedure best suitable for the entire map. In Figure \ref{fig:linmem} we show a comparison of the same region imaged with the two different algorithms.

As stated in Paper I, the multi-scale clean procedure \citep{Cornwell2008} usually results in an imaging improving for fields with extended sources. However we found that an important limitation of this method is the significant requirement of computing resources, at least with the multi-scale implementation provided by \textsc{casa}\footnote{Common Astronomy Software Applications, https://casa.nrao.edu/}. Because of this limit we were not able to use this procedure for the entire map at this moment.

\section{Source extraction and classification}
\label{sec:detcla}
The extraction of point sources in the pilot field was one of the main topic of Paper I. To accomplish that goal, we had used the extraction algorithm by \citet{Franzen2011}. The algorithm extracted sources with dimensions up to 8 times the area of the synthesis beam, but 88 percent of them were consistent with point or very compact sources. A source should be defined a pure point source only when its apparent dimension is identical that of the synthesis beam, that is a deconvolved dimension compatible with 0. However, because of imperfections on the imaging process and because of background noise, the deconvolved dimension of point sources can be greater than 0. On the other hand, the sky at this frequency is full of extragalactic sources, usually characterized by two or three discrete components (namely the two radio lobes and the central regions of the galaxy itself). If these components lie very close to each other the source is only barely resolved and can be easily confused with a slightly extended source. The lower dimension limit for extended sources is then somehow arbitrary. In this work we consider extended sources all those sources whose dimensions are greater than $20\um{arcsec}$ on at least two perpendicular directions. This value corresponds to three times the geometrical mean of the `\textsc{linmos}' map synthesized beam axes and three times the `\textsc{mosmem}' map synthesized beam axis. This should exclude many extragalactic barely resolved sources, since in these cases the source is extended only in one direction (even when there are three components, since they are roughly collinear).

SCORPIO was designed to be an intermediate step between current-technology observations and future surveys. In particular, for the purpose of source extraction and classification, the field is sufficiently small to allow source identification by visual inspection, but it is also sufficiently large for testing and training automated algorithms usually requiring moderately large samples of test objects. In this case, the human-driven visual inspection can be used as a verification check of or a training set for the automated algorithms. In this Section we first visually extract all the extended sources in SCORPIO, as defined in this Section. Then we use the CAESAR algorithm \citep{Riggi2016} to automatically repeat the same process, and compare the results. We eventually focus on the source classification, first collecting all the information in literature and then trying to classify all the extracted sources.

\subsection{Source extraction}
\label{sec:dec}
From a first visual inspection, the extended sources in the SCORPIO field can be roughly classified in two main categories (regardless of their nature): `discrete' sources, i.e. sources with sharp edges and well separated from background, and `diffuse' sources, i.e. low-brightness very-extended sources that gradually fade down to the background level. While sources of the former category are immediately identified, sources from the latter can be easily missed or confused by imaging artifacts. On the other hand, some imaging artifacts can be wrongly considered as true diffuse sources.

Since SCORPIO is a blind survey and a goal is to test source blind detection, the first step was to visually analyze the field and manually extract extended sources on the base of our previous experiences.
Two members of our team examined the whole field independently. At the end, 93 sources were spotted by both the `observers', 2 only by the first one and other 4 only by the second one. This result is encouraging in the sense that different, experienced observers agree in almost all the detections. Since there was not an \textit{a priori} preference for one observer or the other, the final extended sources set was given by the union of the two observers' sets. Taking into account this, the visual inspection led us to identify \ObjTot\ extended sources, \SrcDis\ of the first category and \SrcDif\ of the second. A discussion on the possibility that some real sources were wrongly excluded or some fake sources were considered real is presented in Section \ref{sec:diseac}.

A more meaningful parameter is given by the source dimensions. As we stated in Section \ref{sec:obs}, the maximum LAS that we were able to reach with ATCA is about $4\um{arcmin}$ at $2.1\um{GHz}$. Sources more extended than the LAS are poorly imaged, some of them may have been totally missed for this reason. 
In our sample, 37 sources are more extended than $4\um{arcmin}$. Above this limit we refrain from further quantitative analysis. To overcome this it is necessary to use single-dish observations to supply the zero-baseline. Combining single-dish and interferometer data, we can simultaneously obtain an accurate flux density measurement and the same resolution of interferometer data (e.g. \citealt{Ingallinera2014}). Single-dish observations of the SCORPIO fields are ongoing and we will deal with these sources in a future paper. We also warn that sources slightly below the 4-arcmin limit can be also affected by LAS issues, and they should be analyzed cautiously as well. In Appendix A we report the entire catalogue (Table A1) and a gallery with the images of all the sources.

\subsection{Source extraction using CAESAR}
The source detection process was repeated using the CAESAR algorithm \citep{Riggi2016}, as the first test bench for this algorithm prototype on this kind of data. The algorithm searched for sources in a square region centered at $\alpha=17^\mathrm{h}\ 03^\mathrm{m}\ 34.9^\mathrm{s}$, $\delta=-41^\circ\ 47'\ 47''$ and with a side of $1.1\um{deg}$. In this region 14 extended sources were identified by visual inspection. The algorithm successfully identified 11 of them but failed with three diffuse sources, namely SCO J170105-420531, SCO J170203-415920 and SCO J170300-415534. These three sources are all extremely large, diffuse and characterized by a very low brightness, which can be the reasons why the algorithm failed. However, the main problem was that it extracted roughly 100 sources which seem to be artifacts. We are using the result of this test to tune and refine the algorithm parameters (Riggi et al. \textit{in prep.}).

\subsection{Source classification}
\label{sec:cla}
At the SCORPIO central frequency ($2.1\um{GHz}$), the radio continuum Galactic population is mainly constituted by circumstellar ionized gas nebulae surrounding both young stars (H \textsc{ii} regions) and evolved stars (PNe, SNRs, LBVs, WRs). When they are resolved in radio images, these sources are usually characterized by different morphologies, which can give a first important hint on their nature. If a quantitative analysis is possible, their spectral energy distributions (SEDs) are a significant aid in their classification and characterization too.

To assess how many extended sources are already known, and possibly classified, we used of the most modern and complete catalogues for each kind of source: for confirmed and candidate SNRs \citet{Green2014}, \citet{Whiteoak1996} and Green 2019 \textit{in press.}\footnote{Available at http://www.mrao.cam.ac.uk/surveys/snrs/.}; for PNe \citet{Acker1992}, \citet{Parker2006} and \citet{Miszalski2008}; for \hii s \citet{Anderson2014} (hereafter A14); for WR stars \citet{Rosslowe2015}\footnote{http://pacrowther.staff.shef.ac.uk/WRcat/index.php version 1.21, November 2018.}; no known LBV or LBV candidate lies in the SCORPIO field. From this search we found that among the 99 extended sources, 64 are already known objects (i.e. detected in previous observations). Of these 64 sources, 41 are classified, 14 are candidates of some type of object and 9 are with no classification at all. Of the classified sources there are 39 \hii s, 1 PN and 1 SNR. The candidate known objects include 10 \hii s, 2 PNe and 2 SNRs.

The morphological analysis of different kinds of Galactic sources has proved to be a valuable method for their classification, in particular when 8-$\umu$m images are compared with radio \citep{Ingallinera2016}. While, in Galactic objects, the radio continuum emission always traces the ionized part of the nebula, the infrared (IR) emission at $8\mic{m}$ can be ascribed to different mechanisms and origins. In H \textsc{ii} regions the 8-$\umu$m emission is dominated by molecular lines and bands of polycyclic aromatic hydrocarbons (PAH). These molecules are destroyed by ionizing photons and can survive only beyond the ionization front. Therefore in a typical H \textsc{ii} region the radio emission is wrapped by an 8-$\umu$m emitting shell \citep{Deharveng2010} (see Section \ref{sec:hiicla}). In PNe, instead, the 8-$\umu$m emission arises from a mix of dust continuum and gas lines from highly excited atomic species \citep{Stanghellini2012,Flagey2011}. In a previous work, we showed that for PNe radio and 8-$\umu$m are co-spatial \citep{Ingallinera2016}. Finally evolved massive stars in radio are usually characterized by a radio central object and a surrounding nebula (different nebular morphologies are observed) and there is not a common mutual radio/infrared configuration.

Taking into account our radio observations and IR 8-$\umu$m data from GLIMPSE\footnote{The Galactic Legacy IR Mid-Plane Survey Extraordinaire, conducted withthe InfraRed Array Camera (IRAC) on board the \textit{Spitzer Space Telescope}.} (\citealt{Benjamin2003}; \citealt{Churchwell2009}) we tentatively propose six new \hii s, two new PNe, two new massive star candidates and three new supernova remnant candidates. When the GLIMPSE images were not available we used the \textit{Wide-field Infrared Survey Explorer} (\textit{WISE}) 12-$\umu$m images \citep{Wright2010}. In Table \ref{tab:cla} we report these results and a detailed discussion on each class is presented in the next three Sections.

\begin{table}
\caption{Source classification from the literature and after the discussion presented in this work. No literature source were disputed. Point PNe and WR stars are excluded from this count. All the sources reported in this Tabel are included in Table A1.}
\begin{tabular}{lcc}\hline
Source type & From the & After\\
& literature & this work\\\hline
\hii s or candidates & 49 & 55\\
PNe or candidates & \phantom{4}3 & \phantom{4}5\\
LBV/WR stars or candidates & \phantom{4}0 & \phantom{4}2\\
SNRs or candidates & \phantom{4}3 & \phantom{4}6\\
\hline
\end{tabular}
\label{tab:cla}
\end{table}

\section{Characterization of H \textsc{ii} regions}
\label{sec:hii}
%
%
%
%
%

At the frequency and resolution of SCORPIO, \hii s are likely the most common Galactic radio extended sources. Their study is very important to understand the first phases of the massive star life and their role in the evolution of the interstellar medium (ISM). \hii s have been historically divided into several sub-classes according to their diameter and density (see, for example, \citealt{Kurtz2005}). Even though this division reflects the local physical conditions of an \hii, it is still debated whether each class represents a precise step of the evolution of these objects or not \citep{Kurtz2005b}.

In the Table \ref{tab:hiisub}, we report the canonical sub-class division (e.g. \citealt{Kurtz2005}). The diameter distribution is correlated to the electron density distribution, though the values reported are to be taken as order of magnitudes and the boundaries between the sub-classes are not sharp. This is a clue that the different sub-classes are indicative of different physical conditions. To date an estimate of how many \hii s of each sub-classes can be expected in our Galaxy is missing. An upper limit of the total number of Galactic \hii s can be derived from stellar population models for the distribution of very early-type stars (roughly hotter than B3), the only stars that can produce a detectable \hii. But this is an upper limit since some OB stars can be not associated with an \hii, depending on the ISM conditions, or the same \hii\ can be produced by more than one star. To have at least an estimate of this number, we used the Besan\c{c}on model in order to simulate the expected Galactic population of ionizing stars \citep{Robin2003}. From the model, we got that the number of stars with spectral type not cooler than B2 is $\sim\!2\cdot10^5$ ($\sim\!10^5$ excluding B2). Therefore we can state that the maximum number of Galactic \hii s is of order of $10^5$. Constraining the simulation to the SCORPIO field we obtain an estimated population of $\sim\!4300$ ionizing stars, or $\sim\!1600$ excluding B2. Therefore the total number of \hii s in SCORPIO should not exceed, as order of magnitude, $10^3$.

\begin{table}
\caption{Sub-class division of \hii s \citep{Kurtz2005}.}
\begin{tabular}{lcc}\hline
Sub-class & Size & Electron density\\
& (pc) & (cm$^{-3}$)\\\hline
Hypercompact (HC) & $\lesssim0.03$ & $\gtrsim10^6$\\
Ultracompact (UC) & $\lesssim0.1$ & $\gtrsim10^4$\\
Compact & $\lesssim0.5$ & $\gtrsim5\cdot10^3$\\
Classical & $\sim\!10$ & $\sim\!100$\\
Giant & $\sim\!100$ & $\sim\!30$\\
Supergiant & $>100$ & $\sim\!10$\\
\hline
\end{tabular}
\label{tab:hiisub}
\end{table}

Unfortunately it is complex to foresee how many \hii s we expect to observe for each sub-class, having \hii s proven to be quite heterogeneous objects. For example \citet{Kalcheva2018} compiled a catalogue of UC \hii s from the CORNISH survey \citep{Purcell2008}. The diameter of \hii s in this catalogue ranges from 0.014 to $0.890\um{pc}$, while the electron density from $10^3$ to $10^{4.55}\um{cm}^{-3}$. Reported to the same distance, the flux densities of these catalogued objects would show a variation of more than three orders of magnitude. Due to this extreme variation the derivation of a typical flux density for each sub-class is non-meaningful.

Thus to estimate the completeness of the sample observed with SCORPIO we proceed as follows. We suppose that the logarithms of diameter and emission measure follow a normal distribution and we consider the behaviour of UC \hii s catalogued by \citet{Kalcheva2018} as typical (i.e. all the other sub-classes have the same logarithmic standard deviation). For UC \hii s in this catalogue we would have 
\begin{equation}
\log_{10}\left(\frac{EM}{1\um{pc}\um{cm}^{-6}}\right)=6.5\pm0.5,\ \log_{10}\left(\frac{D}{1\um{pc}}\right)=-0.9\pm0.4,
\end{equation}
where $EM$ is the emission measure and $D$ the object diameter. A region, at a distance $d$, departing $1\sigma$ below both in diameter and emission measure will have a flux density $S_0$. It is possible to show that more than 90 percent of sources located at the same distance will have a flux density greater than $S_0$ (see Appendix B). At $20\um{kpc}$, considering a typical flux density derived from the parameters listed in Table \ref{tab:hiisub}, for a UC \hii\ $S_0\sim\!800\mic{Jy}$, for a compact \hii\ $S_0\sim\!35\um{mJy}$, for a classical \hii\ $S_0\sim\!100\um{mJy}$. Since a distance of $20\um{kpc}$ represents the worst case scenario (being roughly the greatest distance for a Galactic \hii) we can state that for each sub-class more than 90 percent of objects have a flux density above $S_0(20\um{kpc})$.

\begin{figure}
    \includegraphics[width=\columnwidth]{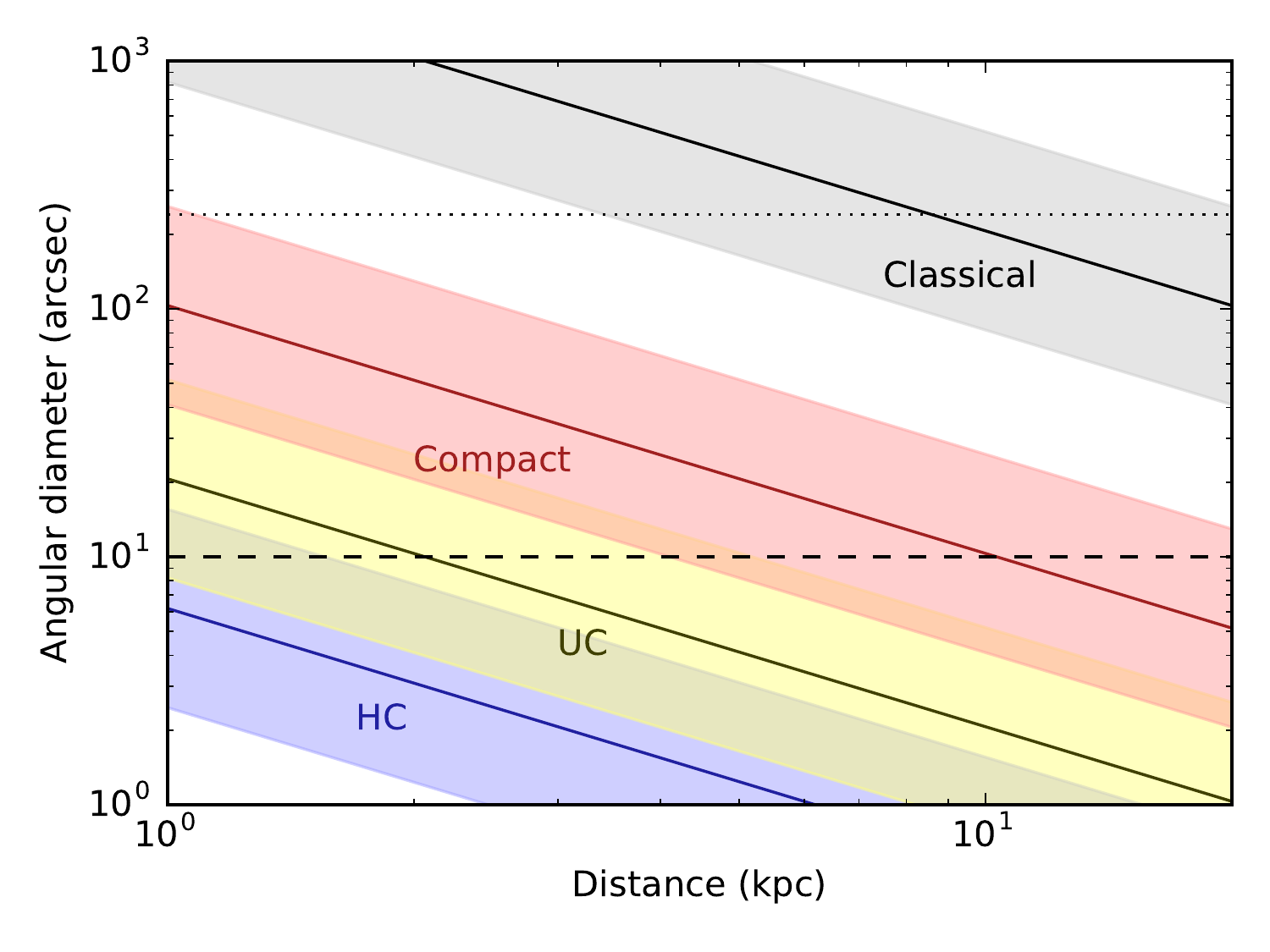}
	\caption{Typical angular diameters for HC, UC, compact and classical \hii s against their distance (solid lines). The shaded regions in the plot represent the $\pm1\sigma$ limit discussed in the Section \ref{sec:hii}, where darker regions correspond to the superposition of two sub-types (i.e. \hii s of different type but with the same diameter).} The dashed line indicates an angular dimension of $10\um{arcsec}$, the minimum for an \hii\ to be resolved in SCORPIO. The dotted line is the SCORPIO LAS ($4\um{arcmin}$).
	\label{fig:hiidiam}
\end{figure}

In Figure \ref{fig:hiidiam} we plot the typical diameters of \hii s against their distance. At the SCORPIO resolution we are able to resolve all the classical \hii s (and those even more extended) and a significant fraction of compact \hii s. Giant and supergiant (and a fraction of classical) \hii s are more extended than the LAS, so they are resolved out in SCORPIO and undergo flux-loss issues. All HC and almost all UC \hii s should appear as point sources. Given these limitations and considering only resolved sources, we focus only on UC, compact and classical \hii s.

With a minimum flux density of $\sim\!35\um{mJy}$ we could safely state that SCORPIO is complete regarding compact \hii s. For UC \hii s the limit flux density of $\sim\!800\mic{Jy}$ is sufficiently high to let us detect ($>5\sigma$) these sources everywhere except in those region on the Galactic plane where deconvolution artifacts lead to a background noise of few hundreds microjansky per beam.

As we said before, we expect to resolve all classical \hii s. For this sub-class we expect a minimum brightness of $\sim250\mic{Jy\ beam}^{-1}$ (taking into account $S_0\sim\!100\um{mJy}$ and an angular diameter of $200\um{arcsec}$), which is still above our detection threshold, however we expect also the nearest and the largest ones could be resolved out. In these cases we should however be able to observe the source, even if its morphology and its flux density could not be recovered properly.

\subsection{Candidate H \textsc{ii} regions}
\label{sec:hiicla}
Among the extended sources in the SCORPIO field, 49 are listed as \hii s or candidates by A14 (see Table A1). Indeed our 49 sources correspond to 48 A14 sources since one of them is associated with two of our extended sources. As we reported in Table \ref{tab:hiiA14}, following the classification of A14, we divided the \hii s in different classes: known (K), radio quiet (Q), group (G) and candidates (C). We consider all the first three categories as confirmed \hii s.

\begin{figure}
    \includegraphics[width=\columnwidth]{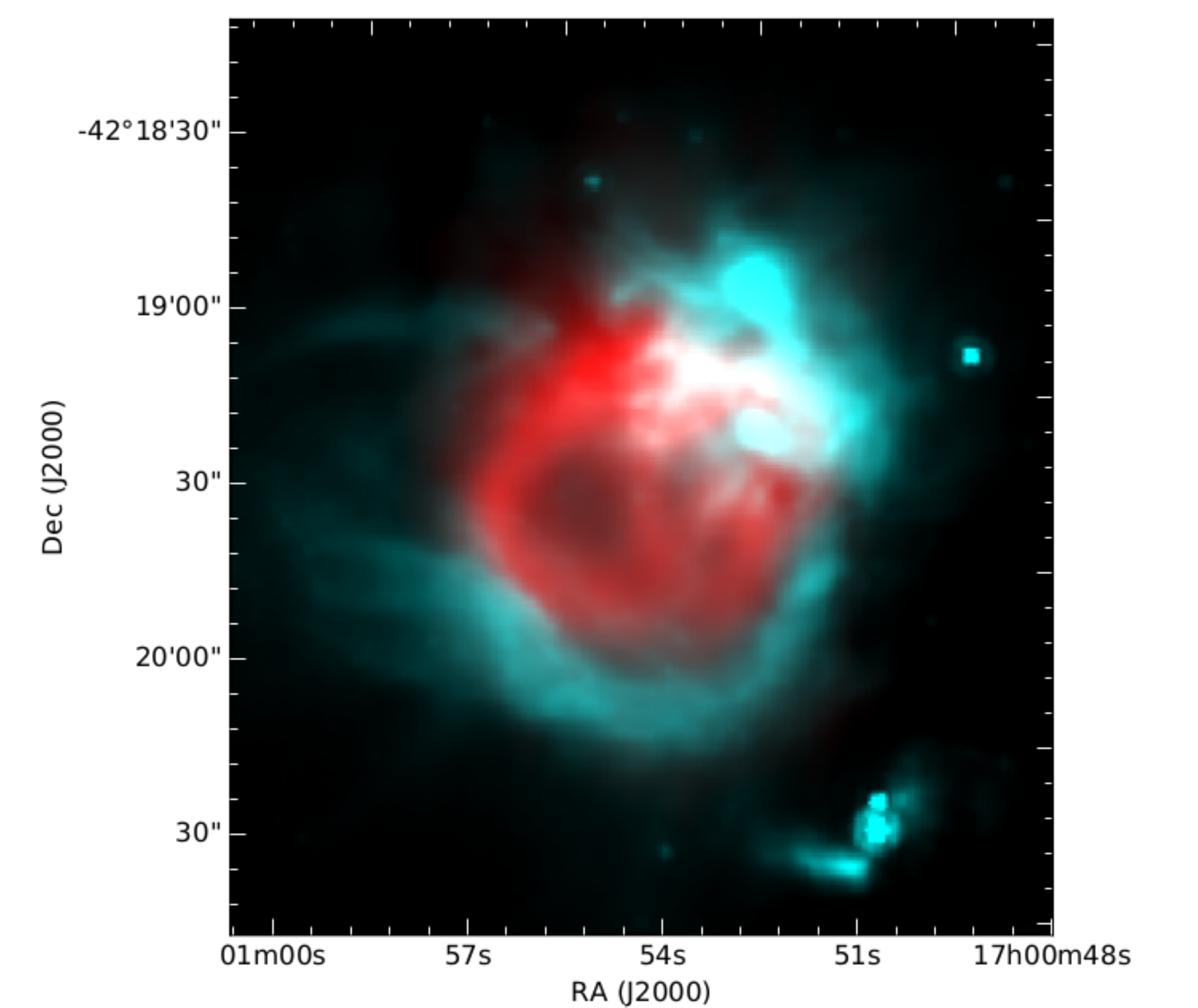}
	\caption{Two-color image of the known \hii\ SCO J170054-421917 (in red our radio data, in cyan GLIMPSE at $8\mic{m}$). It is possible to appreciate how the prominent 8-$\umu$m emission completely wraps the radio.}
	\label{fig:anello_rgb}
\end{figure}

To verify the A14 classification, for each \hii\ we checked the mutual position between radio and 8-$\umu$m emission, using the GLIMPSE images. As we mentioned in Section \ref{sec:cla}, we considered that the radio continuum emission supports the \hii\ classification if the 8-$\umu$m emission wrapped it, otherwise the \hii\ nature is questioned. A clear example is shown in Figure \ref{fig:anello_rgb}. Among the 48 \hii s detected we were able to confirm the A14 classification for 47 of them, while only for SCO J165538-415101, classified by A14 as candidate, and for SCO J165903-424139, classified by A14 as radio-quiet, our analysis was inconclusive and further investigation is required.

Given the potentiality of this analysis, we applied this method also to all the other 35 non-classified extended sources in SCORPIO. We found that other six sources show a mutual radio/8-$\umu$m disposition typical of \hii s and therefore we propose all these sources as \hii s, for a total of 55 \hii s or \hii\ candidates in SCORPIO. This supports the hypothesis that, at this frequency and with these sensitivity and resolution, the \hii s represent the large majority of all extended Galactic sources.

The total number of \hii s catalogued by A14 in the SCORPIO field is 165, of which 48 detected by our observations as extended sources, and discussed above, 25 associated with a point source in our maps, 14 are close to other A14 regions associated to the same extended source in this work, and 78 not detected at all. Taking into account what we said in Section \ref{sec:hii} about angular extensions (see also Figure \ref{fig:hiidiam}), the point \hii s are likely to be UC \hii s. A spectral index study on these sources is ongoing (Trigilio et al. \textit{in prep.}). The case of undetected \hii s is much more interesting. Among these 64 are classified by A14 as `radio-quiet', 8 as candidate and 6 as group. In table \ref{tab:hiiA14} we summarize this result.

\begin{table}
\caption{Detection of A14 \hii s. For each type defined by A14 (`C': candidate, `G': group, `K': known, `Q': radio quiet),} we report the number of \hii s associated with SCORPIO extended sources, with SCORPIO point sources, with other \hii s in A14 and with no association in SCORPIO.
\begin{tabular}{cccccc}\hline
Type & Ass. with & Ass. with & Ass. with & No radio & Total\\
& extended & point & other A14 & detection\\
& sources & sources & \hii s\\\hline
C & 10 & 4 & 2 & 8 & 24\\
G & 1 & 0 & 10 & 6 & 17\\
K & 28 & 3 & 0 & 0 & 31\\
Q & 9 & 18 & 2 & 64 & 93\\
Total & 48 & 25 & 14 & 78 & 165\\
\hline
\end{tabular}
\label{tab:hiiA14}
\end{table}

Many of the `radio-quiet' \hii s are not detected in our observations likely because their emission is below the sensitivity limit, and in fact 29 of them are detected as extended or point source. Of the other non-detected \hii s, five `group' and four `candidate' lie very close to the edge where the background noise is very high. The other non-detected `group' and one non-detected `candidate' are reported by A14 as much larger than our LAS, and this can be the reason why we miss them. The remaining three non-detected `candidate' are located on the Galactic plane and may be associated with the very extended sources SCO J170020-42343 and SCO J170818-41093.

Considering also point \hii s, we can state that the SCORPIO field harbors at least 171 \hii s or candidates, of which six proposed for the first time in this work. It is difficult to estimate how many \hii s we are not detecting (besides the 78 detected by A14 and not by our observations) or we are detecting but failing to correctly identify as \hii s. However, the total number of \hii s in the SCORPIO field is, at least as order of magnitude, what we could expect from theoretical considerations. In the previous Section we stated that the total number of ionizing stars in SCORPIO should be of order of $10^3$. \citet{Tremblin2014} used the \hii\ Region Discovery Survey (HRDS; \citealt{Anderson2011}) to conduct statistical studies on the age of 119 \hii s. The age of these sources mostly range between 0 and $5\um{Myr}$, with the majority around $1.5\um{Myr}$. Given the typical lifetime of each class (e.g. \citealt{Keto2002}), about 70 percent of them are classical \hii s and around 25 percent are UC or compact (so the other types represent a marginal fraction of the total). The most abundant ionizing stars are early B stars, characterized by a lifetime $\sim\!20\um{Myr}$. So a typical ionizing star spends only $10-20$ percent of its life forming an \hii. So we expect that the total number of \hii s will be a factor $0.1-0.2$ the total number of ionizing stars. For SCORPIO this translates into around $10^2$, which is what we found as order of magnitude. Therefore SCORPIO likely has a high degree of completeness with respect to these three sub-classes.


\section{Discussion on planetary nebulae}
\label{sec:pn}
PNe are among the most studied Galactic objects, since they play an important role in many astrophysical processes. Because they represent the possible last stages of low- and intermediate-mass stars (zero-age main-sequence mass between 0.8 and $8\um{M}_{\sun}$), a better comprehension of PNe would result in a significant improvement in understanding the stellar evolution and the material and energy exchange with the ISM. It has also been suggested that the different morphologies exhibited by PNe may be shaped, among other factors, by magnetic fields \citep{Sabin2007}, so their study has implications on fundamental plasma physics too. Deep radio observations are going to supply a great contribution to the study of Galactic PNe, in particular in detecting missing PNe expected from stellar evolution models (e.g. \citealt{Moe2006}). Toward the Galactic plane, in fact, the main discovery probe, i.e. the H $\alpha$ emission, may be hampered by the absorption, by a very high and complex background and by the confusion noise. Also IR observations may suffer from these problems, mainly from the very high background (especially for $\lambda\gtrsim8\mic{m}$). Furthermore, 24-$\umu$m observations, which is the band where PNe are expected to show the peak emission, currently suffer from a limited resolution that prevents, in many cases, morphological considerations \citep{Ingallinera2016}. Finally, radio observations can be used to estimate the distance of a resolved PN \citep{Ingallinera2014,Ingallinera2016}. In this case it is also possible to derive important physical parameters like the dimensions, the electron density and the ionized mass, which are fundamental to constrain stellar evolution models.

\subsection{Planetary Nebulae in SCORPIO}
The SCORPIO field harbours ten PNe and PN candidates known from literature and reported in the MASH catalogue \citep{Parker2006}, in the Strasbourg-ESO Catalogue of Galactic Planetary Nebulae \citep{Acker1992}, in \citet{Acker2016} and in the HASH catalogue \citep{Parker2016}. Three PNe are catalogued in MASH, one of them is classified as `true' PN, the other two as `likely' PNe (we will generically refer to them as `PN candidates'), and we detect all of them as extended sources. The other three catalogues contain other 7 PNe or candidates, but five of them are not resolved in SCORPIO, one is located in a very confused region so nothing can be said about its morphology and one (PN HaTr 5) is not detected at all. Two extended sources in SCORPIO show radio and infrared morphologies that closely recall the classical elliptical PNe. A summary of all PNe is reported in Table \ref{tab:pne}.

\begin{table}
\caption{List of all known, candidate and proposed PNe in SCORPIO. References: $^a$\citet{Acker1992}, $^b$\citet{Parker2016}, $^c$this work, $^d$\citet{Parker2006}, $^e$\citet{Acker2016}.} 
\begin{tabular}{lccc}\hline
Name & Confirmed? & Resolved in & Ref.\\
& & SCORPIO?\\\hline
H 1-3 & yes & no & $a$\\
H 1-5 & yes & no & $a$\\
IC 4637 & yes & no & $a$\\
IRAS 16515-4050 & yes & no & $b$\\
MGE 343.6641+00.9584 & no & yes & $c$\\
MGE 344.1648+00.2733 & no & yes & $c$\\
MPA1654-4041 & no & yes & $d$\\
PHR1654-4143 & no & yes & $d$\\
PN G343.5+01.2 & yes & yes & $d$\\
PN G344.3+02.1 & yes & no & $e$\\
PN G344.4+02.1 & yes & no & $e$\\
PN HaTr 5 & no & - & $a$\\ 
\hline
\end{tabular}
\label{tab:pne}
\end{table}

The `true' PN is PN G343.5+01.2 (SCO J165355-414357), a very compact source whose dimensions, as reported in MASH, are $15\times11\um{arcsec}^2$. Given these dimensions, our radio images show only a barely resolved source. Anyway the radio emission is spatially coincident with the 12-$\umu$m emission from \textit{WISE} and with the H $\alpha$ emission. No more information about the morphology of the nebula can be extracted, nevertheless, in accordance with its classification, we can exclude a radio/IR configuration similar to \hii s. This source was included in the catalogue released in Paper I with a spectral index, derived exploiting the ATCA wide band, $\alpha=-0.07\pm0.06$, with $S\propto\nu^\alpha$, compatible with a free-free emission \citep{Cavallaro2018}.

For the MASH candidate PHR1654-4143 (SCO J165451-414345) radio and 12-$\umu$m emission are coincident with H $\alpha$, in agreement with a possible PN nature and excluding the possibility of an \hii. 

For the second MASH candidate MPA1654-4041 (SCO J165443-404145), the radio emission, though the source is barely resolved, strictly recall the same elongated structure observed in H~$\alpha$ and at $8\mic{m}$, where signature of a central object are also present. Indeed 8-$\umu$m image suggests a bipolar shape rather than elliptical as catalogued in MASH. Its morphology could be a hint of a very young PN or even a proto-PN. Indeed we find a spectral index of $\alpha=-0.6\pm0.1$, which can be compatible with this hypothesis (e.g. \citealt{Cerrigone2017}). Both this source and PN G343.5+01.2 are much less extended than the LAS across the entire observing band, so their spectral index should not be affected by any inaccurate flux density measurement.

As stated before, other two sources in SCORPIO have an elliptical radio morphology typical of a PN. The first one, SCO J170019-415042, is an already known radio source, detected by the MGPS (MGPS J170019-415039). At $24\mic{m}$ it appears as a `disk' nebula as catalogued by \citet{Mizuno2010} (MGE 344.1648+00.2733). It is also detected at $8\mic{m}$, with a morphology similar to radio, but it immersed in a very high and confusing background. Probably for the same reason, the nebula is not detected in H $\alpha$. The second, SCO J165546-414839, has no previous radio records but it is clearly detected at $24\mic{m}$ as a ring nebula (MGE 343.6641+00.9584). It is not detected at $8\mic{m}$ nor in H $\alpha$, very likely because of its intrinsic faintness (brightness peak in radio $0.6\um{mJy\ beam}^{-1}$, that is about $0.5\um{MJy\ sr}^{-1}$). The non-detection in H $\alpha$ may be also due to the very bright background in that direction. These two sources are the only MIPSGAL bubbles as catalogued by \citet{Mizuno2010} in the SCORPIO field. If both these candidates were PNe the idea that most of the MIPSGAL bubbles are PNe would be further supported \citep{Ingallinera2016}.

\subsection{Searching for missing PNe}
\label{sec:misspne}
It is widely accepted that the number of known Galactic PNe ($\sim\!3000$) is less than the real total number of Galactic PNe by about one order of magnitude (with estimates of the total number varying between 6600 and $46\,000$; \citealt{Sabin2014}). Likely, this discrepancy derives from the limited sensitivity of the current surveys in the Galactic plane, especially in the optical and IR as stated in the previous section. In this Section we evaluate which contribution to solve this problem radio surveys like SCORPIO or EMU can provide in future, taking into account not only the sensitivity limit but also the possibility to resolve the sources. In fact, even if a PN is detected in these radio surveys, it could be difficult to identify it as a PN if it appears as a point source, since, for example, their SED can be confused with that of other different kind of sources.

\begin{figure}
	\includegraphics[width=\columnwidth]{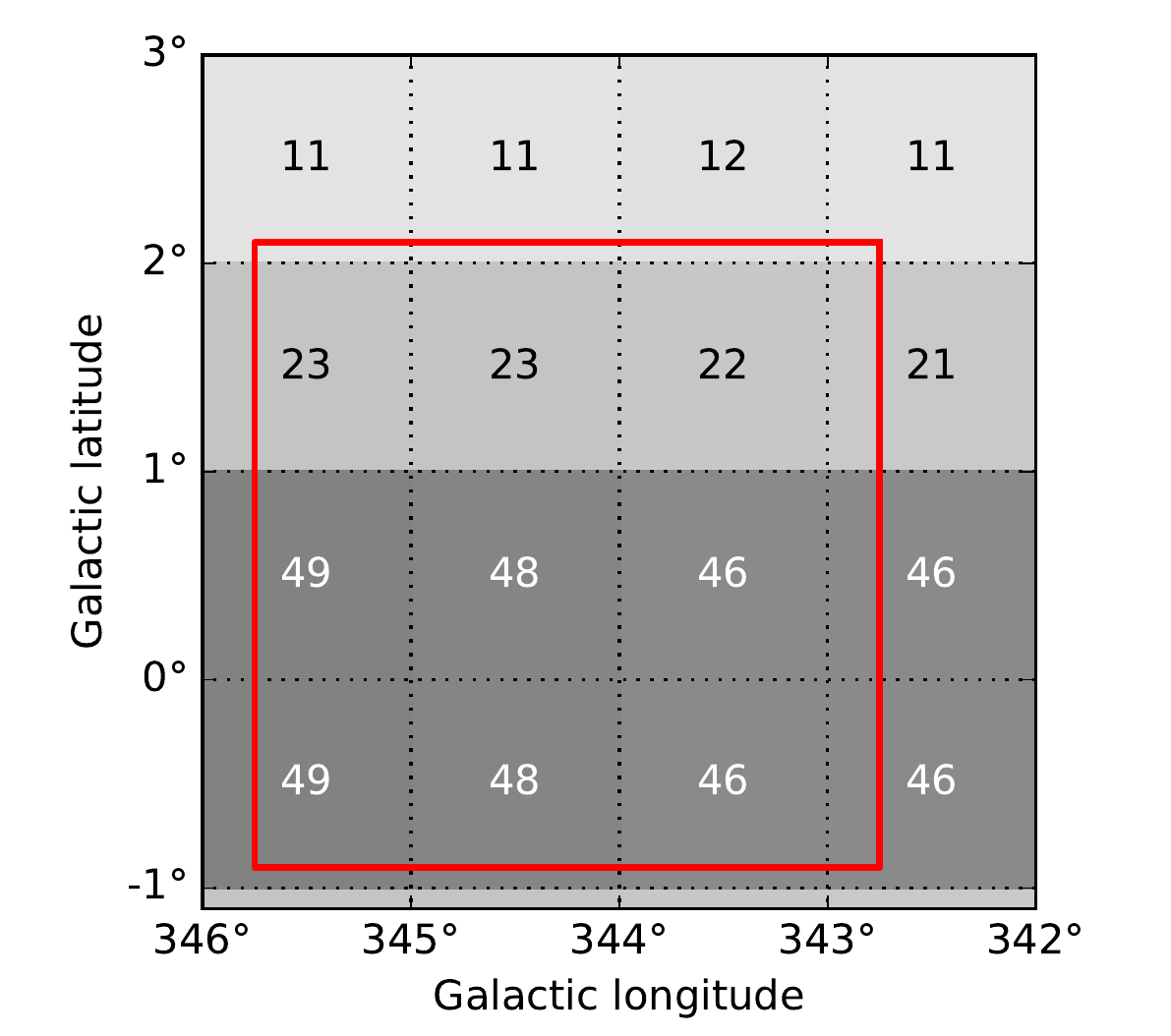}
	\caption{Number of PNe per square degree toward the SCORPIO region (delimited by the thick line), assuming a total number of $20\,000$ Galactic PNe and an exponential distribution.}
	\label{fig:densPN}
\end{figure}

First of all we estimate how many PNe we should expect in the SCORPIO field. To this goal many different distribution models can be used, as well as the total number of Galactic PNe. In this work we are merely interested to an order of magnitude estimate so the choice of the model is not critical. We decide to use an exponential disk
\begin{equation}
n=n_0\exp\left(-\frac{\rho}{\rho_0}\right)\exp\left(-\frac{z}{z_0}\right),
\end{equation}
where, in a cylindrical reference frame, $n$ is the PN volume density, $\rho$ is the radial coordinate (from the Galactic centre), $z$ is the height from the Galactic plane, $n_0$ is chosen so that the total number of PNe is $20\,000$, $\rho_0=3.5\um{kpc}$ \citep{Gilmore1989} and $z_0=250\um{pc}$ \citep{Zijlstra1991}. In Figure \ref{fig:densPN} we report the density of PNe found using this model toward the SCORPIO field.

\begin{figure}
	\includegraphics[width=\columnwidth]{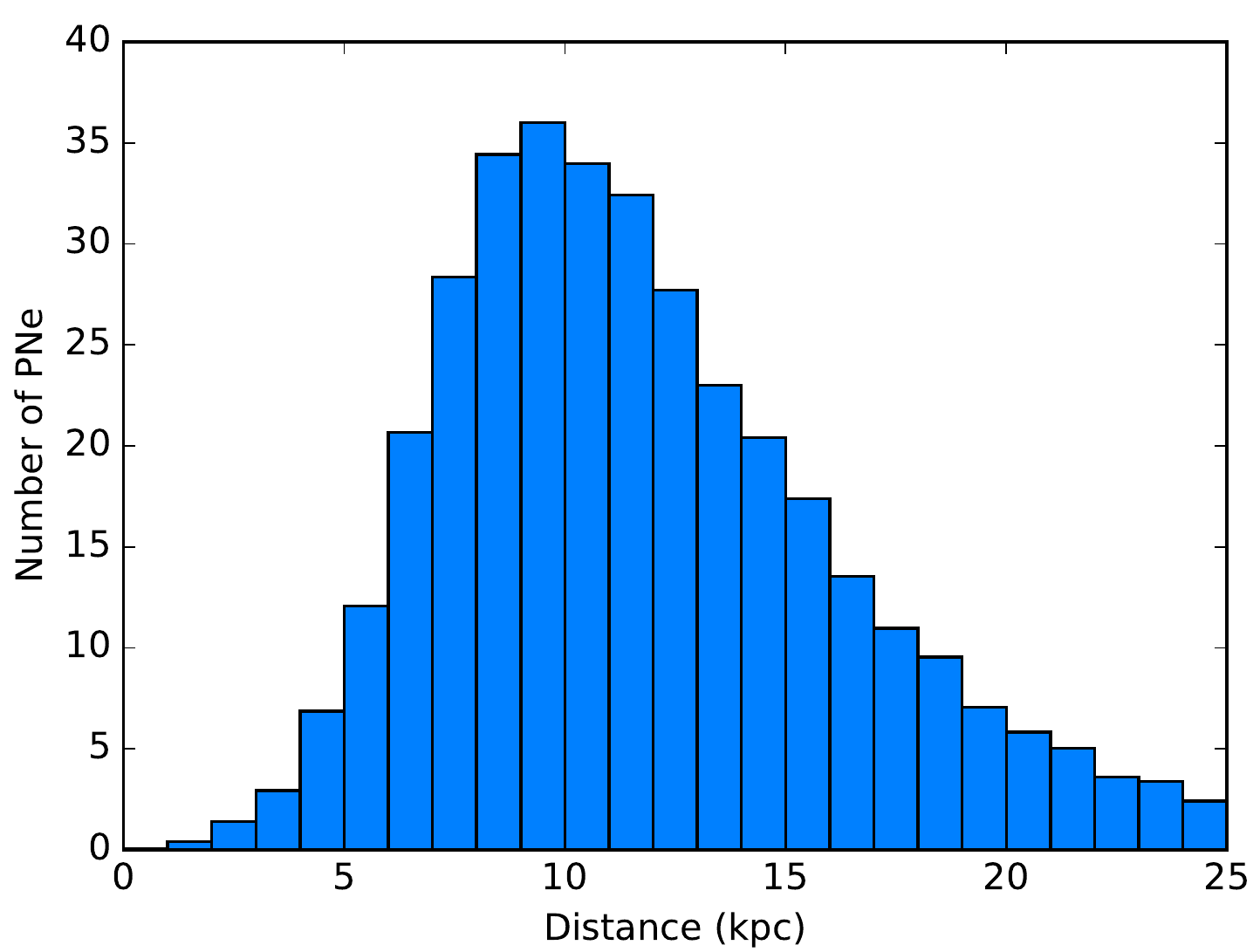}
	\caption{Predicted distribution of the distance from the observer of the PNe in the SCORPIO field.}
	\label{fig:hist_dist}
\end{figure}

The model we used predicts about 350 Galactic PNe in the SCORPIO field (results from the simulation range from 341 to 359). Given the conditions stated above, about sensitivity and resolution, we do not expect to detect or identify all of them. In order to understand if our model is in agreement with what we observe, we first use the model to build the distribution of the distances and flux densities. The distance distribution can be directly obtained from the model and is reported in Figure \ref{fig:hist_dist}.

For the distribution of the flux densities, we need an idealized emission model for PNe. We suppose that a PN is a spherical nebula that emits in radio only via free-free mechanisms. In this case the flux density of the nebula at a given distance and at a given frequency is a function of: the nebula radius ($R$), the electron density ($n_e$), the electron temperature ($T_e$), the filling factor ($\epsilon$) and its chemical composition. It is also known that the ionized mass of a PN is related to its radius (e.g. \citealt{Kwok2007}). For our purpose we use the relation by \citet{Zhang1995}:
\begin{equation}
\log\left(\frac{M_\mathrm{ion}}{\mathrm{M}_{\sun}}\right)=1.3077\log\left(\frac{R}{1\um{pc}}\right)+0.3644.
\label{eq:mass}
\end{equation}
From the ionized mass we derive the mean electron density as:
\begin{equation}
n_e=\frac{3M_\mathrm{ion}}{4\pi m_pR^3},
\label{eq:ne}
\end{equation}
where $m_p$ is the proton mass. We assume an electron temperature $T_e=10^4\um{K}$, a filling factor $\epsilon=0.6$. For the chemical composition we only take into account the Helium and we apply a corrective factor $Y=1.258$, assuming $\mathrm{He/H}=0.11$ and $\mathrm{He^+/He}=0.5$ (see \citealt{Kwok2007}). Hence we derive the mean brightness of the source as
\begin{equation}
I(\nu)=\epsilon Y B_{\mathrm{bb}}(\nu,T_e)(1-e^{-\tau}),
\label{eq:I}
\end{equation}
where $B_{\mathrm{bb}}(\nu,T_e)$ is the brightness of a black-body and $\tau=\tau(\nu,T_e,n_e)$ is the optical depth calculated after \citet{Oster1961}.

We now assume that the radii of PNe are distributed as in the sample used by \citet{Jacob2013}. With this assumption and with equation (\ref{eq:I}) we derive the distribution of the expected flux densities. We obtain that in the SCORPIO field there could be about 150 PNe with a flux density above $2\um{mJy}$. We consider a PN `fully resolved', that is resolved enough to allow a morphological classification, if its diameter is greater than $30\um{arcsec}$ and we consider such a nebula detectable if its flux density is above $5\sigma\sqrt{\Omega}$, where $\sigma=100\mic{Jy\ beam}^{-1}$ is the map standard deviation and $\Omega$ is the solid angle subtended by the source expressed as number of synthesis beams. Under these conditions, our model predicts about 10 detectable fully resolved PNe. Considering all the assumption, this number is in agreement with the five fully resolved PNe truly observed.

If this estimate is correct, a survey like EMU could potentially observe more than half of the total expected Galactic PNe. However, the main limitation does not appear to be the sensitivity of the survey but the resolution, since many potentially detectable PNe would not be recognized.







\section{Evolved massive stars and supernova remnants}
\label{sec:mass}

\subsection{Luminous blue variable and Wolf--Rayet stars}
Massive star evolution is still heavily debated. It is not clear which evolutionary phases these stars undergo as a function of their initial mass. This, in fact, depends on the mass-loss history of each star, which can be peculiar and influenced by surrounding events (e.g. a companion star; \citealt{Sana2012}). Based on observational evidences and theoretical or numerical arguments, different authors have suggested that some blue supergiants pass through a LBV phase, before to evolve as WR stars or directly explode as SN (e.g. \citealt{Langer1994}; \citealt{Kotak2006}; \citealt{Trundle2008}; \citealt{Umana2012}; \citealt{Groh2013}). During the LBV phase the stars undergo several mass-loss episodes, whose violence can sometimes resemble a SN explosion (e.g. $\eta$ Car; \citealt{Davidson2012}).

The SCORPIO field does not harbour any known LBV or LBV candidate, but four known WR stars, namely HD 151932, HR 6265, HR 6272 and SFZ12 1168-91L. The first three are all detected as point sources, while the last one is close to the edge of the field in a very noisy region and we are not able to detect it. Since none of them is an extended source they are not considered in this work.

Both LBV and WR stars are characterized by strong stellar winds and are usually hot enough to ionize their circumstellar envelope (CSE). Several radio observations of confirmed LBVs in our Galaxy and in the Large Magellanic Cloud have shown these two signatures of ionized ejecta \citep{Duncan2002,Umana2005,Agliozzo2012,Buemi2017}. In \citet{Ingallinera2016}, we discussed how these two features have a peculiar signature in radio continuum observations when compared to other sources of radio emission (e.g. PNe or SNRs). In particular these stars are usually associated with a diffuse radio nebula, whose geometry can vary from circular to strongly bipolar, with a prominent central object well-detected at centimetre wavelength. The radio nebula traces the emission of the ionized CSE while the radio emission of the central object is due to the stellar wind. This appearance well distinguishes massive evolved stars from other kind of Galactic objects like \hii s or PNe. We also showed that these objects can have a differentiated spectral index (like HR Car; \citealt{Buemi2017}). Following \citet{Ingallinera2016} we searched for massive evolved candidates in the SCORPIO field, looking for a radio nebula with a central object.

\begin{figure*}
	\includegraphics[height=6cm]{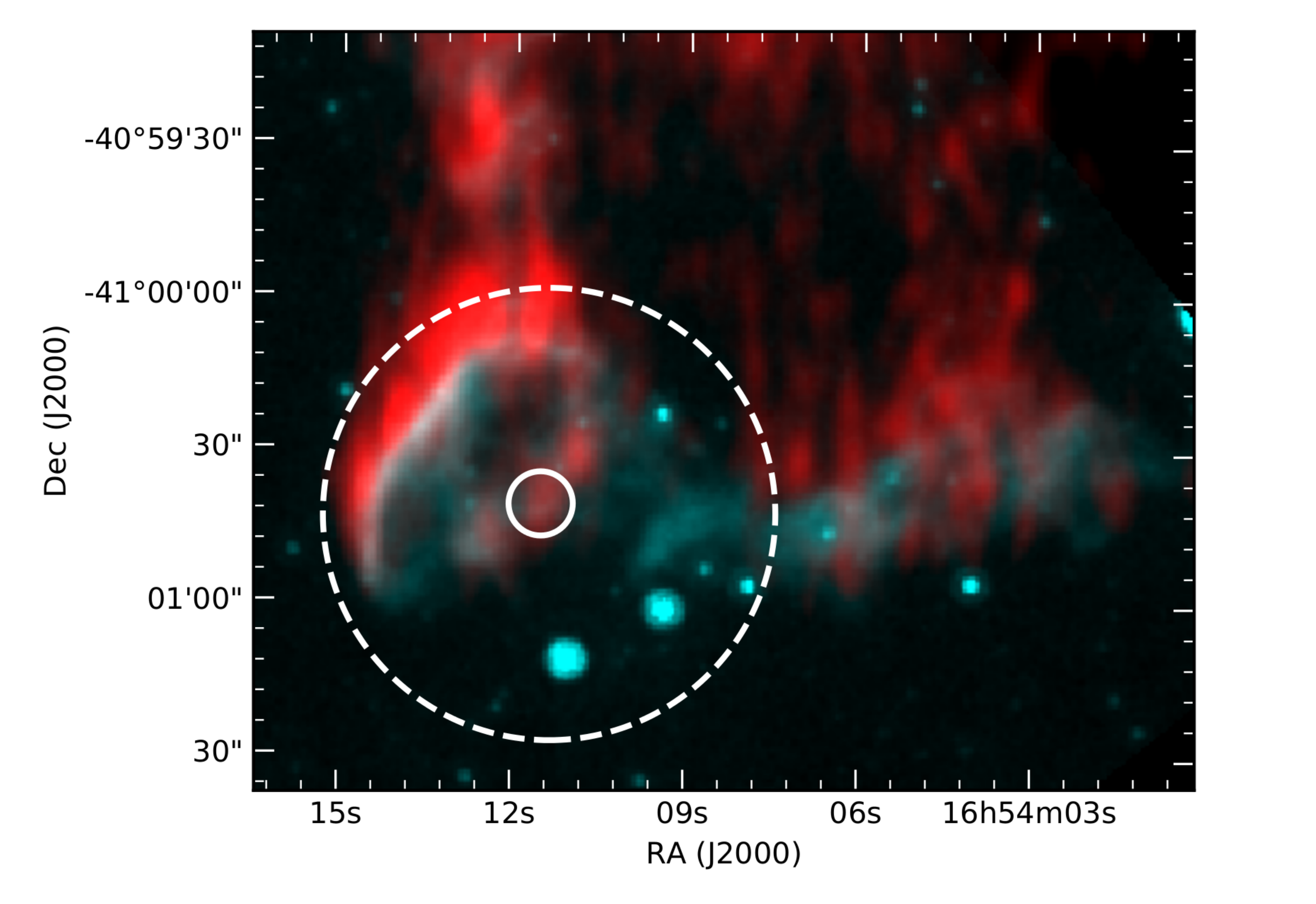}
    \includegraphics[height=6cm]{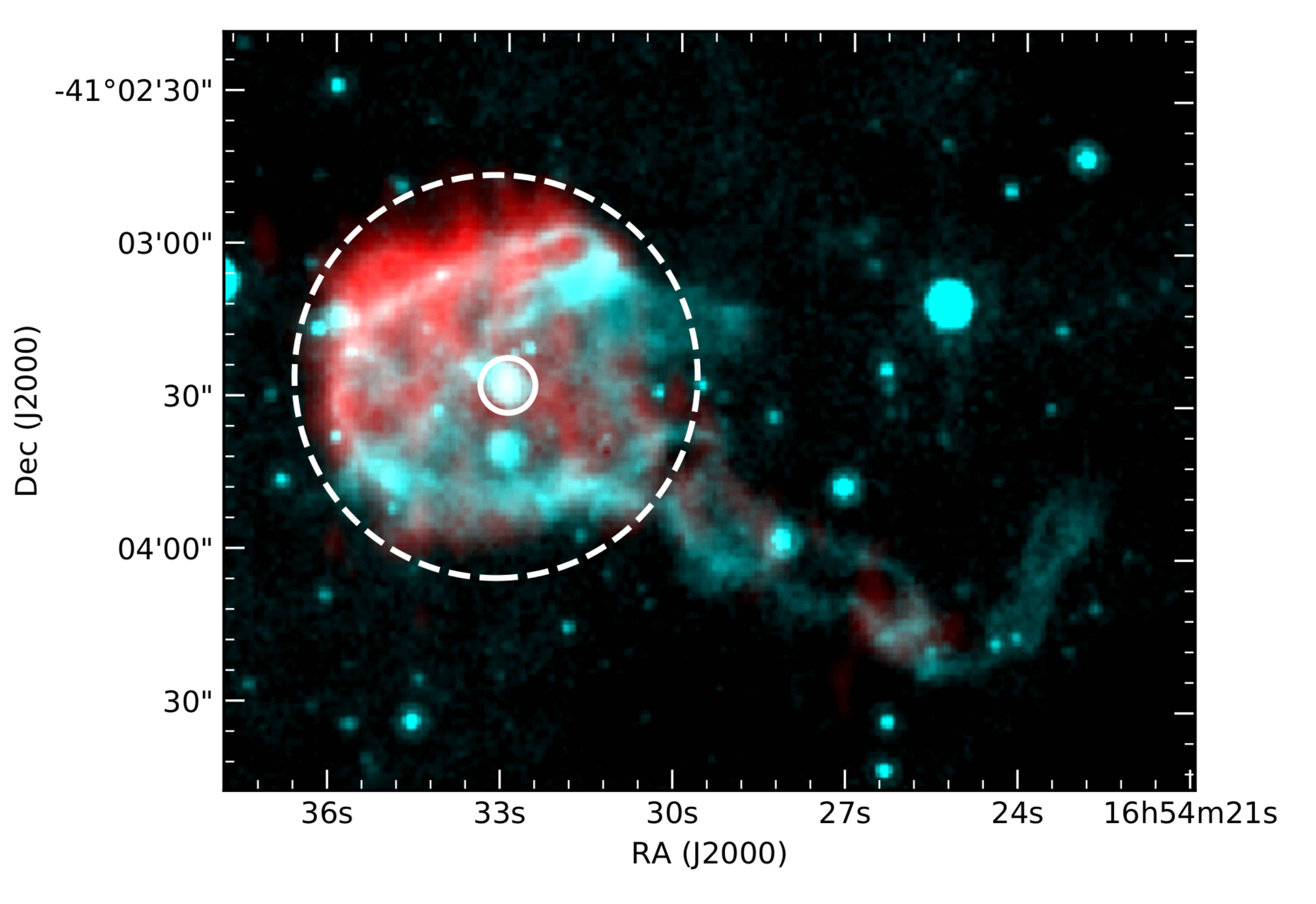}
	\caption{Two-color images of the LBV/WR star candidates SCO J165412-410032 (left) and SCO J165433-410319 (right; in red our radio data, in cyan GLIMPSE at $8\mic{m}$). The dashed circle highlights the external shell, while the solid circle the central object.}
	\label{fig:lbvc}
\end{figure*}

Using this method we found two sources in SCORPIO satisfying this requirement: SCO J165412-410032 and SCO J165433-410319. Neither of these two sources had been previously reported in literature. Remarkably similar, in our maps they appear as a point source surrounded by a roughly circular nebula. Comparing with IR data, we can see how the central object and the nebula are clearly detected also at this wavelength (confirming \citealt{Ingallinera2016}; for a discussion on the mid-IR properties of the ejecta see \citealt{Egan2002}). But in the IR images, in both the sources, a double-helix structure clearly departs from the centre of the nebulae (see Figure \ref{fig:lbvc}). Likely these structures could be jets modelled by the interaction with a companion star. Other examples of blue supergiant or WR stars with jets in the literature are WR102c \citep{Lau2016} and MWC 137 \citep{Mehner2016}.

According to what we have discussed about these two objects, we propose them as LBV or WR star candidates. A spectroscopic analysis on the central object should be able to corroborate and further refine this classification.

\subsection{Supernova remnants}
\label{sec:snr}
Supernova explosions produce a SNR that is in general easily detectable in radio. Theoretical estimates predict that in our Galaxy about 1000 SNRs should be present, but to date we have been able to detect only about 300 of them \citep{Green2014}. The great majority of these SNRs is detected in radio. It is therefore reasonable that current and future radio surveys are going to allow us not only to better characterize known SNRs but also to discover missing ones. Indeed, these surveys overcomes many of the limits that have possibly hampered SNR characterization or further detections, combining high resolution, high sensitivity and a several-arcminute LAS.

The SCORPIO field hosts just one known SNR (SNR G344.7-0.1, listed in Table A1 as SCO J170358-414314; \citealt{Green2014}) and two candidates from the literature (MSC G345.1+0.2 and MSC G345.1-0.2, listed in Table A1 as SCO J170346-410659 and SCO J170514-412549; \citealt{Whiteoak1996}). SNR G344.7-0.1 is likely a composite-type SNR that has been extensively studied in radio, IR, X-rays and $\gamma$-rays \citep{Giacani2011}. In the region of this remnant our map is very similar, in resolution and sensitivity, to that published by \citet{Giacani2011} at $1.4\um{GHz}$. Unfortunately the extension of this SNR (about $9\um{arcmin}$ in diameter) is significantly greater than our LAS ($\sim\!4\um{arcmin}$), preventing us to perform any quantitative study.

The SNR candidate MSC G345.1+0.2 \citep{Whiteoak1996} appears as a ring-shaped source, incomplete on the north-west part, with a diameter of $\sim\!11\um{arcmin}$ and a thickness of $\sim\!2.5\um{arcmin}$. There is not a clear evidence of IR emission associated with the radio nebula, neither in GLIMPSE or in MIPSGAL images, supporting its classification as a SNR.

On the other hand, the SNR candidate MSC G345.1-0.2 \citep{Whiteoak1996} is a disk source, with a diameter of $\sim\!6.5\um{arcmin}$. Also in this case, there is not a clear evidence of IR emission associated from GLIMPSE and MIPSGAL images. As for MSC G345.1+0.2, this seems to corroborate the hypothesis that this object could be a SNR.

\begin{figure*}
	\includegraphics[height=4.75cm]{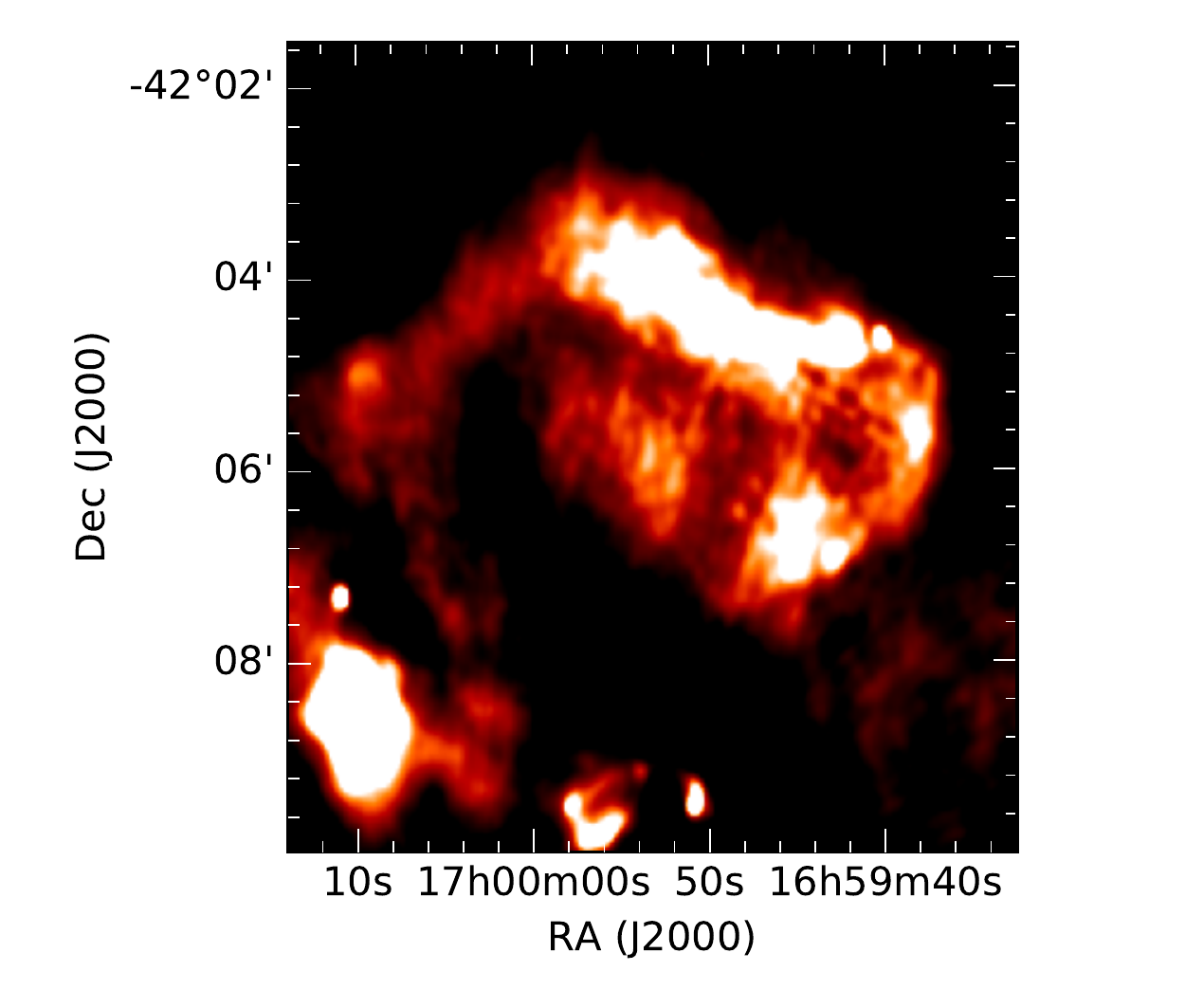}\hspace{3mm}
    \includegraphics[height=4.75cm]{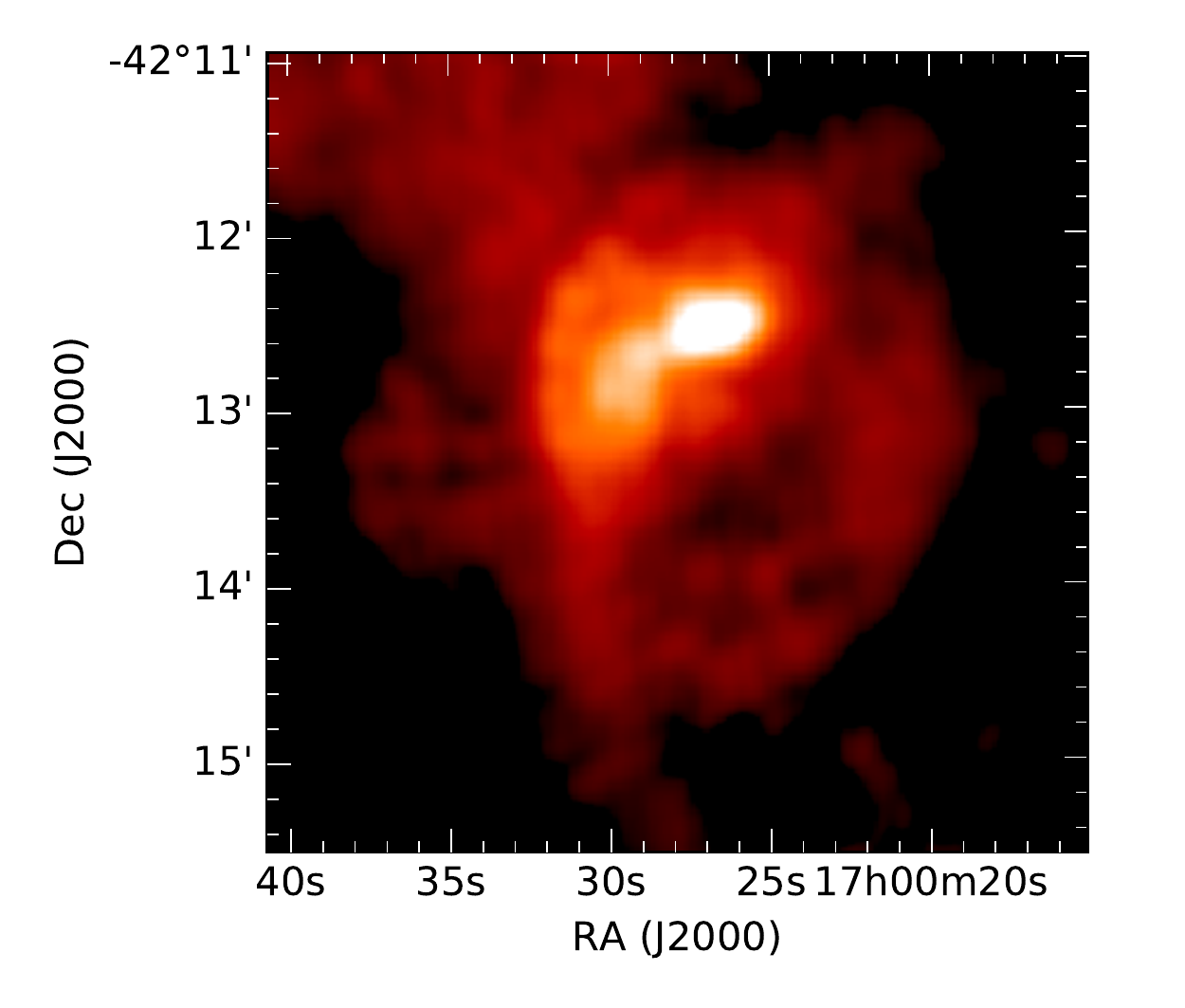}\hspace{3mm}
    \includegraphics[height=4.75cm]{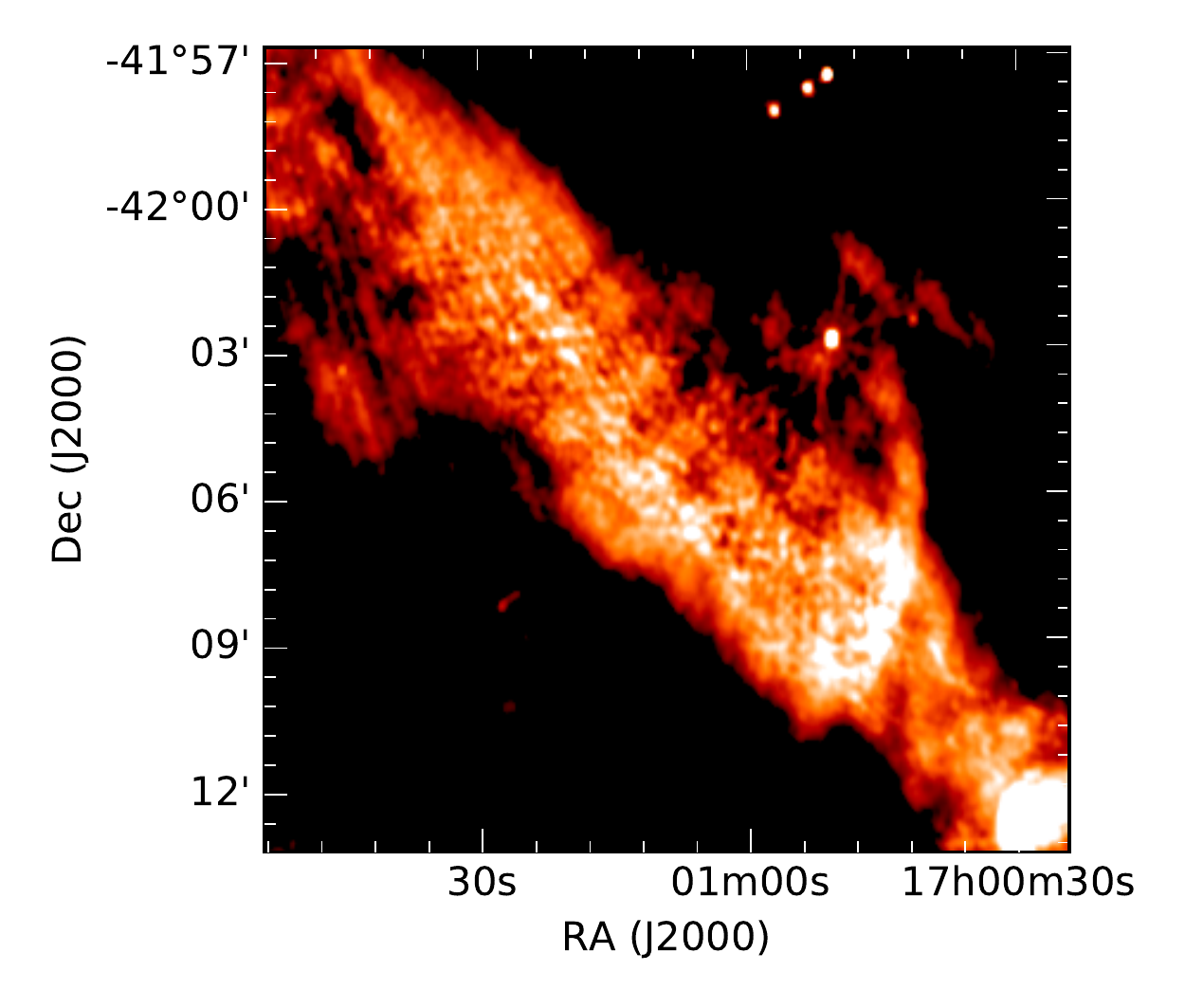}
	\caption{Radio images of the extended sources SCO J165948-420527 (left), SCO J170029-421309 (center) and SCO J170105-420531 (right), proposed in this work as SNR candidates.}
	\label{fig:snrp}
\end{figure*}

The lack or weakness of the IR counterpart to the radio nebula together with the overall roundish morphology can be used as a criterion to confirm the nature of such candidates as SNRs or to search further new SNRs. We have to warn that no one of these two criteria is strictly necessary, since some SNRs do show a significant IR emission (though weakness is a strong signature; e.g. \citealt{Ingallinera2014a}) or can divert from a round shape (e.g. plerions). Following these stringent criteria, at least three sources, reported in Figure \ref{fig:snrp}, deserve a special attention. The first, SCO J165948-420527, is located in a region where different discrete sources (mainly \hii s but also unclassified objects) are present too. It appears as a diffuse source with an extension of about $7\um{arcmin}$. There is not a correlation with IR, except in a couple of compact regions. This source can possibly be a real SNR with some \hii s prospectively coincident with it or it is just a denser region of the ISM.

The second, SCO J170029-421309, is located exactly on the Galactic plane ($b\sim0^\circ$). It appears as a roundish source with a diameter $\sim\!4\um{arcmin}$ and an enhanced brightness toward the centre. The presence of the GDE makes its identification particularly difficult and uncertain. Lying on the Galactic plane, the IR maps at 8 and $24\mic{m}$ are dominated by the dust thermal continuum and therefore it is very difficult to search for a IR counterpart. However there is not a striking evidence of a possible IR source associated with the radio nebula.

The third, SCO J170105-420531, is also located on the Galactic plane but it shows a prominent arc extending above the Galactic plane in the north-west part of the nebula. If that arc were a complete circle the source diameter would be $\sim\!13\um{arcmin}$. In the IR images there is no evidence of the arc, while nothing can be said on the region lying toward $b=0^\circ$. Our hypothesis is that the arc can be reasonably part of a previously unobserved SNR, possibly superimposed to the GDE.

All the SNRs or candidate discussed in this Section are characterized by angular extensions greater than or approximately equal to the LAS. As we said, unfortunately this prevents us from any further quantitative analysis on these objects. Nevertheless, with these observations we were able to corroborate the proposed nature of the two SNR candidates from literature and to propose other three SNR candidates.

\section{Discussion}
\label{sec:dis}

\subsection{Results and limits on source extraction and classification}
\label{sec:diseac}
One of the main goals of current and future surveys is the extraction of all the detected sources. But if on one hand it is fundamental to be able to miss as least sources as possible, on the other, especially with interferometer data, it is crucial to avoid misinterpretations of imaging artifacts as real sources (a `false positive' problem). For compact sources or for sources with well-defined edges (those that we called `discrete' in Section \ref{sec:dec}) this task is relatively easy for an expert, well-trained astronomer. For `diffuse' sources, sometimes also characterized by a very low brightness, the task is much harder and some ambiguity may rise if we base only on radio data. For this reason, as explained in Section \ref{sec:dec}, we decided that the manual source extraction would have been performed by two independent astronomers, in order to subsequently cross-check their results. Both the astronomers were able to identify 93 over the 99 final sources. Though this result is reassuring, we are not able to surely asses the reality (whether they are true sources or artifacts) of two sources (SCO J170203-423142 and SCO J170414-424328) even after further analysis and comparison with IR. These two sources both appear as `diffuse' sources with a dimension of $\sim\!1.1$ and $\sim\!5.0\um{arcmin}$ respectively. Beyond the fact that they do not present any IR counterpart, they seem detected only in the `\textsc{linmos}' map and not in the `\textsc{mosmem}' map. For the first source a possible cause for the discrepancy may be the different background noise level of the two maps in that region (respectively $\mathrm{rms}\sim\!30$ and $\mathrm{rms}\sim\!100\mic{Jy\ beam}^{-1}$) with respect to a mean brightness of the source around $\sim\!100\mic{Jy\ beam}^{-1}$ (comparable to `\textsc{mosmem}' map rms). The second source is located close to the edge of the field in a region with severe artifacts in the `\textsc{linmos}' map and high noise in the `\textsc{mosmem}' map. Therefore doubts on the reality of both sources remain. The same doubts had been raised also for other sources in our catalogue but typically the comparison with IR clarified their nature (see also Section \ref{sec:gde}).

One of the main issues affecting our maps is represented by imaging artifacts. Just like in any other interferometric images the origins of artifacts are to be searched in: unflagged RFI and bad data; calibration errors; imperfect $uv$-plane coverage; chosen deconvolution algorithm. The editing and calibration procedure represented an important phase in our data reduction. Even if we cannot exclude residual errors in this process, these should be negligible with respect to those deriving from the imperfect coverage of the $uv$ plane. In order to improve the $uv$-plane coverage we added data from observations in compact configurations, obtaining almost the best one achievable with ATCA. Despite this, the $uv$-plane coverage is not fully satisfactory yet and, obviously, the center of the $uv$ plane is not, and it cannot be, covered at all, since we are dealing with interferometric observations. This has implications on the LAS of the observations that, considering also that short baselines are the most affected by the flagging process, is not greater than $\sim\!4\um{arcmin}$ (see Section \ref{sec:obs}). As we said, there are 37 sources more extended than $\sim\!4\um{arcmin}$ and, if fully recoverable, the GDE would be a continuous source crossing the whole map. This causes the presence of the classical artifacts due to the lack of short-spacing information in the $uv$ plane, like the negative bowl-shaped regions around extended sources. These artifacts cannot be further mitigated only with ATCA data. The problem gets worse for regions close to the map edges where some (even external) extended sources cannot be CLEANed, producing artifacts. This issue hampers also the possibility to carry out a quantitative analysis on very extended sources. Finally both the chosen deconvolution and imaging algorithms and their tuning parameters influence the quality of the resulting image and, among the two we used in this work, the maximum entropy method gives the best result for extended sources. Other algorithms, or their combinations, may be more efficient but in this case were impractical from a computational point of view.

The dimensions of the SCORPIO field are particularly suitable to start the automation of some fundamental tasks, like Galactic source extraction and classification. SCORPIO is in fact sufficiently small that these tasks can still be performed manually, but also sufficiently large to allow us to develop and test automated tools. We had started to develop the automated extraction tool CAESAR exploiting the SCORPIO field data \citep{Riggi2016}. The need for this new tool was motivated by the fact that most of the tools developed and tested in radio astronomy are mainly oriented to point source extraction. Therefore these algorithms do not satisfactorily fit the necessities of Galactic studies, where a significant amount of observed sources are extended (provided a resolution of $10\um{arcsec}$). The number of sources missed by the algorithm is relatively low but it still needs further fine tuning. Currently, the hardest obstacle is represented by the imaging artifacts, which makes tricky even the manual extraction, both in searching for real sources but also in reaching the theoretical noise level in critical regions of the map. For this reason, beside progressing in the algorithm capabilities, we think that more efforts have to be made in improving imaging algorithms.

With these data we obtained an important confirmation of the radio surveys capability in source classification. Indeed one of the most promising potentiality of new high-sensitivity high-resolution surveys is the possibility to derive hints on the nature of detected sources starting from their radio morphology. The other classification diagnostic tool used in radio, the spectral index study, despite being applicable also to non-resolved sources, shows degenerate values for different types of Galactic sources. Objects like \hii s, PNe and evolved massive stars may be characterized by a very similar SED in the centimetre wavelengths. Taking into account what we showed in \citet{Ingallinera2016}, in this work we were able to propose two new PN candidates, two new LBV/WR candidates and three new SNR candidates, based only on their radio morphology. The availability of IR images, above all around the band 8-$12\mic{m}$, allows us to further expand the classification possibility. 
Beside this, IR images allow us to classify sources whose radio morphology is not sufficient for this purpose.
This is the case of \hii s, where the mutual position of radio and 8- or 12-$\umu$m emission permitted us to corroborate the nature of 47 over the 49 \hii s reported in literature (9 over the 10 reported as candidate in literature) and to propose 6 new \hii s. Considering that the survey carried out with \textit{Spitzer} and \textit{WISE} already provide an almost complete coverage of the Galactic plane in the 8/12-$\umu$m band with a resolution of few arcseconds, the impact of surveys like EMU to the study of the various Galactic object populations will be tremendous. At this moment the main limitation seems to be the resolution, given that no morphological analysis can be made on point sources. Therefore this work also remarks the necessity of the next-generation instruments operating in these bands (radio and 8-$12\mic{m}$) with an even higher resolution, such as SKA and the \textit{James Webb Space Telescope}. In particular with this work we stress the importance of planning Galactic observations for these two instruments, but also to address and optimize their data exploitation and their synergy.

\subsection{The Galactic diffuse emission}
\label{sec:gde}
The study of the GDE can be dated back to the beginning of the radio astronomy when, because also of the low observing frequency and the low resolution, they observed an indistinct diffuse emission permeating the entire Galactic plane (e.g. \citealt{Bridle1967}). At frequencies from $\sim\!100\um{MHz}$ to $\sim\!1\um{GHz}$ the Galactic emission is dominated by a non-thermal component with a spectral index $\alpha$ ranging from $-0.5$ to $-0.7$ (e.g. \citealt{Rogers2008}; \citealt{Platania2003}). This is likely a synchrotron emission coming from free relativistic electron interacting with the Galactic magnetic field. It is clear that this is only a very partial and incomplete picture, and it cannot be representative of the Galactic emission behaviour across the radio band. At a resolution around $1\um{deg}$ at a frequency $\sim\!1\um{GHz}$ or higher the contribution from Galactic thermal sources, which becomes dominant at freqeuncies above $\sim\!10\um{GHz}$, results in a flattening of the global spectral index \citep{Dickinson2003}. Furthermore, in principle we cannot exclude that different components of the ISM could have different spectral behaviours. Building a comprehensive model of the Galactic emission has become an important task in the last decades, not only to achieve a satisfactory description of the Galactic dynamics and structure, but also for cosmological studies regarding the cosmic microwave background emission (CMB; see \citealt{de2008}). In this last context it is fundamental to model the Galactic emission to subtract the proper foreground and recover an accurate picture of the CMB. In order to obtain this result it is necessary to take into account the different emitting components that concur to the total emission. In a recent work, \citet{Zheng2017} built a model of the Galactic emission in a wide frequency range, from $10\um{MHz}$ to $5\um{THz}$. At the SCORPIO central frequency, $2.1\um{GHz}$, the model has a resolution $\sim\!1\um{deg}$ and predicts that the synchrotron emission is only a factor 2-3 higher than the free-free emission. It is reasonable to think that most of the free-free contribution comes from \hii s, as suggested also by our SCORPIO images where \hii s are the most abundant kind of Galactic source at this frequency. It is however difficult to say which fraction of the free-free emission is represented by \hii s.

Our SCORPIO maps harbors 29 diffuse sources characterized, as we stated in Section \ref{sec:detcla}, by not well-defined borders and fading out smoothly to the background. Eight of these sources are indeed \hii s known from the literature. It is difficult to establish the nature of the other sources, especially for those very extended and weak, also because many of them are more extended than the LAS and are heavily affected by imaging artifacts. As we said in Section \ref{sec:diseac}, for two of them we are still in doubt about their reality, since potentially some imaging artifacts may create impostor fake sources. In general, to establish the reality of a source, we took into account those radio data at a frequency similar to that of SCORPIO that, though with a lower resolution, have a greater LAS (e.g. the data that we are collecting with ASKAP and presented in Umana et al. \textit{in prep.} or the data from the Molonglo Galactic Plane Survey) or contain single-dish observations (e.g. Parkes data that we are collecting within the EMU collaboration or from the Southern Galactic Plane Survey; see Section \ref{sec:diseac}). The detection of the same source with a morphology compatible in different maps is a strong indication that we are observing a real source rather than imaging artifacts.

\begin{figure}
	\includegraphics[width=\columnwidth]{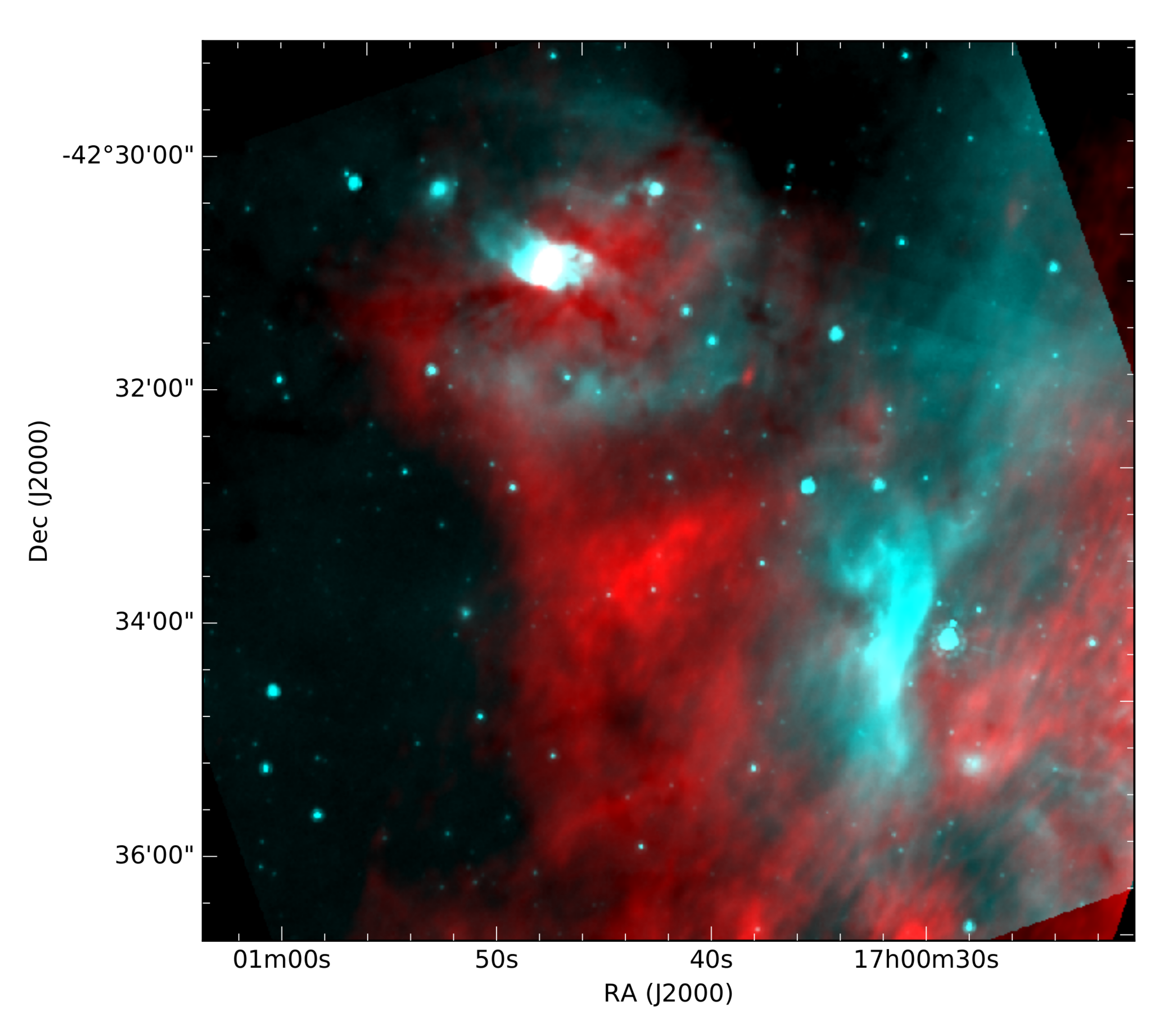}
	\caption{Two-color image of the known \hii\ SCO J170049-423153 (in red our radio data, in cyan GLIMPSE at $8\mic{m}$). This is an example of a diffuse radio source which is a known \hii s and can be used as a prototype for this kind of objects.}
	\label{fig:gdehii}
\end{figure}

\begin{figure}
	\includegraphics[width=\columnwidth]{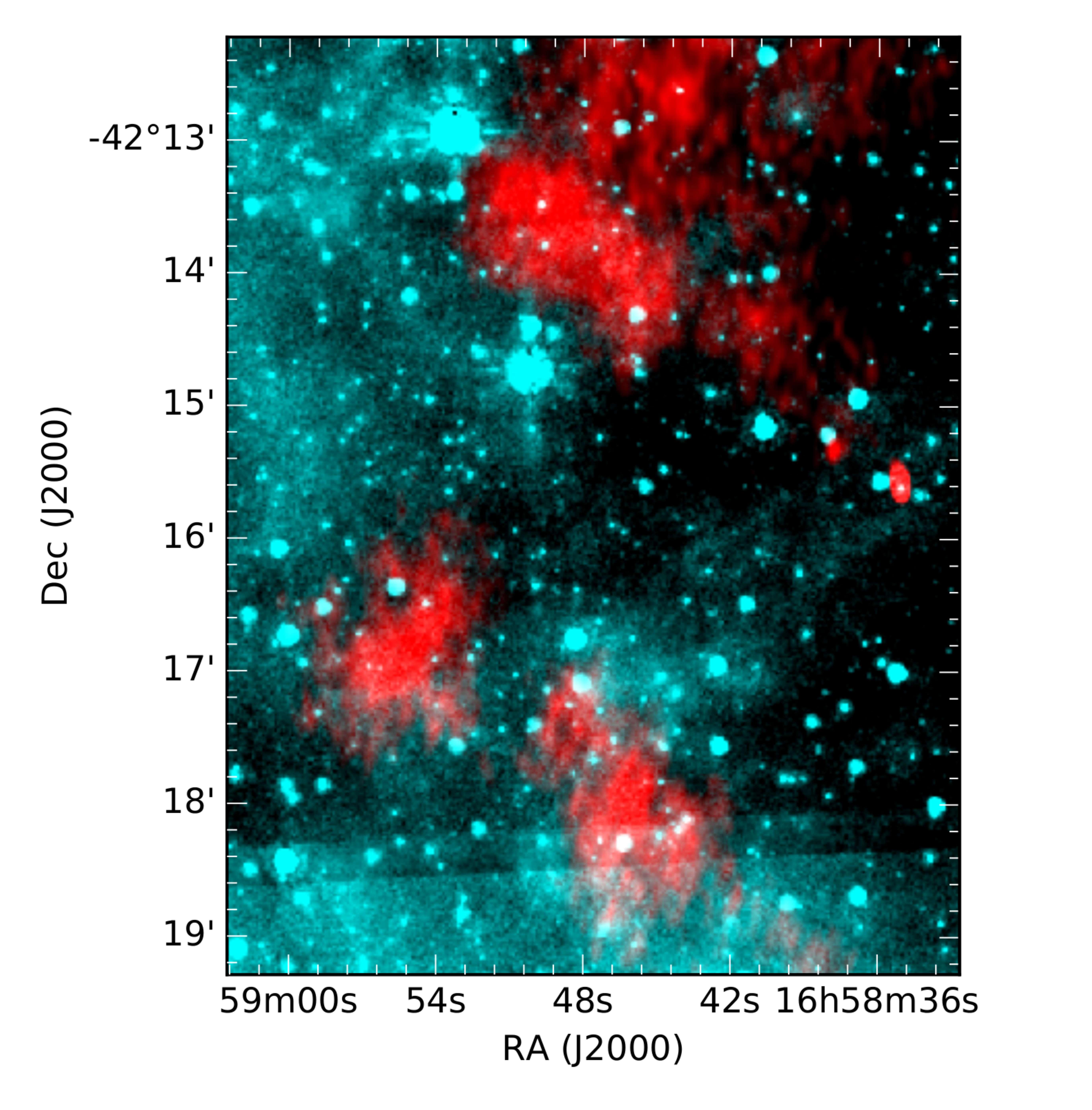}
	\caption{Two-color image of the extended sources SCO J165843-421308, SCO J165844-421814 and SCO J165855-421638 (in red our radio data, in cyan GLIMPSE at $8\mic{m}$). All these three sources are new detection, and we classified them as diffuse sources. While the top sources has an anticorrelating counterpart in the IR, the bottom ones do not.}
	\label{fig:gdenew}
\end{figure}

As we said, it is noteworthy that seven of the sources that we classified as diffuse and that we might have considered simply GDE are indeed classified as \hii s by A14 (an example is reported in Figure \ref{fig:gdehii}). At least other three diffuse sources in SCORPIO without a classification in literature seem to show hints of an \hii\ nature (see Figure \ref{fig:gdenew}). In our radio maps they appear as a diffuse emission but the comparison with the IR shows that some of them are, partially, surrounded by a 8-$\umu$m emission in a similar way of discrete \hii s. This feature is not seen in other sources classified as GDE in our catalogue. A possible explanation is that we are observing very extended \hii s. Given the low brightness, these \hii s should be characterized by a low density. These sources fit the characteristics of the so-called extended low-density warm ionized medium \citep{Petuchowski1993}. As described by \citet{Gordon2002}, since the ISM is not homogeneous but irregular and clumpy, some of the ionizing photons of hot stars may escape the immediately surrounding medium, spread across the ISM and may ionize low density regions of the ISM with extension up to hundreds of parsecs. It is therefore possible that some `hot' stars (in the sense of Section \ref{sec:hii}), despite being born in dense clouds, are now in a extremely-low density medium and their UV photons are able to ionize distant clouds. If in some direction these resulting \hii s are still photon bounded then it is possible to expect the 8-$\umu$m emission from PAH beyond the ionization front. Otherwise, if the resulting \hii s are density bounded, we do not expect to detect the 8-$\umu$m emission. Likely both scenarios may apply simultaneously for these extended sources.

This, very simple, model is able to explain the appearance of the radio and 8-$\umu$m emission of some of our diffuse sources, and explains why the emission in these two bands is anti-coincident. Our observations seem to corroborate that the $\mathrm{H}^+$ in the ISM can exist in different `phases' and that at least a part of it is ionized by the UV photons of hot stars and emits in radio by a free-free mechanism. In this case we could expect that these diffuse sources present a spectral index $\alpha\sim-0.1$, rather than $-0.5$ or $-0.7$. Unfortunately our observations do not allow a quantitative analysis of these sources and probably the only effective way to probe their spectral behaviour is to use single-dish observations, especially at higher frequencies. In this case the interferometric observations would be used as a guide to recognize and remove discrete sources not resolved with a single-dish or to focus the observations in areas devoid of prominent discrete sources. Another test for this hypothesis is to search for radio recombination lines toward diffuse sources, a signature of \hii s and historically the first method to detect these very extended low-density sources  \citep{Gordon2002}.


\section{Summary and conclusions}
\label{sec:con}

The main scientific goal of the ATCA observations of SCORPIO was the creation of a catalogue of different populations of Galactic radio sources. In Paper I we released the point source catalogue related to the `pilot field', while the whole field point source catalogue is in preparation. In \citet{Cavallaro2018}, we showed how it is statistically possible to separate the Galactic point source population from the ubiquitous extragalactic contaminants. In this work we showed how it is possible to classify most of the extended Galactic sources, distinguishing between the different classes also with the IR images aid. In this direction, we compiled a catalogue of these extended sources, highlighting which difficulties we encountered and what can be said about newly discovered sources.

The second scientific goal was the study of CSE as a fundamental step for better understanding the stellar and Galaxy evolution. In this work, taking also into account \citet{Ingallinera2016}, we showed that we are able to distinguish CSE surrounding young stars (\hii s) from CSE surrounding both low- and high-mass evolved stars. Though the characterization individual sources is not one of the goals of this article (we reserve this task for future works) we showed some examples where our radio observations proved extremely useful (for example in proposing new PNe and SNRs). The main obstacle to their full characterization is represented by the LAS achievable with ATCA, that does not allow us to proper image some of our sources. This limitation is sensibly mitigated by instruments like ASKAP and MeerKAT, which are characterized by a greater LAS at their operating frequencies. Despite this issue, we provided statistical considerations on different kind of Galactic sources with a CSE (\hii s and PNe), showing in some case how it is possible to derive population considerations. These results can be used to have a reliable estimate of the potential of future surveys to predict what kind of Galactic objects we could detect.

In the Sections \ref{sec:hiicla}, \ref{sec:misspne} and \ref{sec:snr} we estimated, whenever possible, the completeness of SCORPIO. In fact we discussed about the problem of the missing objects, that is about the discrepancy, for different kind of objects, between the expected population predicted by models and the number of objects actually observed. This problem is particularly significant for PNe and SNRs, for which it is estimated that only respectively 15 and 30 percent have been observed. Their detection has a double importance. First, their discovery represents a test bench for population models, usually based not only on extrapolations of what already observed but also on stellar evolution models: a more complete census provides a quantitative support to these hypotheses. Second, the objects belonging to the same class are characterized by physical parameters with a different degree of homogeneity and these parameters may depend on the initial mass of the star, on its chemical composition, on its evolutionary history and on the surrounding environment. All these parameters are critical inputs to stellar evolution models and have a strong influence on determining, for example, the very last phase of the stellar life. Incomplete observations of a class of objects risk to be biased toward particular parameter values and may therefore give us a distorted picture of these phenomena. This second reason is even more relevant for those objects like LBV stars for which the number of expected and observed objects is very low and so every new discovery can significantly improve our knowledge.

The next step toward the classification and the characterization of the extracted extended sources will be the automation of these tasks. For example, as we discussed in Section \ref{sec:cla}, the different morphology in radio and $8\mic{m}$ can be used as an immediate tool to discriminate between H \textsc{ii} regions and PNe. The development of such an algorithm is currently ongoing and different approaches, from analytic to machine learning, are being investigated. It is worth noting that the development of automated algorithm both in source extraction and characterization is critical when the large volume of data from future large surveys, like EMU, will render a human-based visual inspection on such huge fields totally impractical.

The main technical goal of SCORPIO was to find a data reduction strategy to satisfactorily image a patch of the Galactic plane at the target resolution and sensitivity. SCORPIO was in fact the first deep Galactic survey to map a relatively wide region with a resolution of $10\um{arcsec}$. The encountered problems mainly derived from the presence of complex extended and diffuse sources that notoriously makes the interferometric imaging process difficult. As we discussed in detail in Section \ref{sec:red}, we used two different approaches to get to the final maps (what we called the `\textsc{linmos}' and the `\textsc{mosmem}' maps). Our conclusion was that both approaches could be considered valid, although optimized for different aspects. The first is characterized by a lower noise in areas devoid of extended sources, while the second better performs in imaging such sources. Both the final maps present the classical artifacts due to extended sources, deriving from the lack of data at very short baselines and then not fully removable. Other artifacts encountered during the data reduction were instead ascribable to non-flagged data and calibration errors and were subsequently corrected.

The two main problems that affect the extended sources are that their image is rendered imprecisely and that their flux density is underestimated by an unknown quantity. The artifacts corrupt also the areas of the map close to extended sources, where the map noise (better estimated by the standard deviation) increases sensibly up to two orders of magnitude and is no more random. This hampers the point source extraction too, since in those areas the sensitivity is effectively reduced. Our mitigation strategy was based on the improvement of the $uv$ plane at the best of the ATCA capabilities (Section \ref{sec:obs}). Once we reached the instrument limits, we verified that in the neighbourhood of very extended sources or close to the Galactic plane the background noise did not decrease when the integration time further increased and therefore the artifacts represent our depth limit on the Galactic plane with ATCA. This statement does not exclude that it is possible to carry out targeted Galactic observations with a lower noise level, but it could not be possible everywhere in the Galaxy or at least it could not be applicable to a large survey. In light of EMU, our observations indicate that also ASKAP may reach a noise limit on the Galactic plane much higher than the EMU target sensitivity of $10\mic{Jy\ beam}^{-1}$ and that this limit cannot be pushed by increasing the integration time. The estimation of this limit is unfeasible \textit{a priori}, and the experimental estimation is ongoing thanks to the ASKAP early-science observations of SCORPIO between 800 and $1800\um{MHz}$ (Umana et al. \textit{in prep.}). For ASKAP, we could also rely on a much better $uv$-plane coverage that should significantly mitigate this issue.

A further improvement of the $uv$ plane, both for ATCA and ASKAP, to overcome this limit is possible only combining the observations with single-dish data. Observations with the Parkes telescope of the SCORPIO field are in progress (Section \ref{sec:gde}), with the explicit aim to supply zero-baselines to interferometric data, in particular to allow a quantitative analysis also for very extended sources and mitigate imaging artifacts. The combination of intereferometric and single-dish data may be excessively demanding from the computational point of view and these observations are investigating this point too. The technical problems arising from rapidly variable sources (especially when datasets separated by several months are joined together) are still to be tested.

The results we are obtaining from the ATCA observations of the SCORPIO field are already impacting the ASKAP early science. Indeed the ATCA observations have allowed us to deeply know this patch of the sky, providing us information on the sources harboured and on the technical problems to face also with ASKAP. For example the ASKAP data reduction of Galactic data has been inspired also by what we learned so far, with inputs supplied for fine tuning the reduction pipeline for the Galactic plane. Finally the SCORPIO data with ATCA are serving as validation for the ASKAP data.




The wide-context purpose of this work was to offer a panoramic view of the potentialities of the next-generation radio Galactic surveys through our pathfinder observations. This work does not exhaust the project itself but represents, along with the Paper I, an introduction for possible future works. We already started our work on ASKAP data, taking into account the progresses made with the SCORPIO data reduction and analysis. We plan to use the catalogue presented in this work to conduct individual and detailed studies on the most interesting sources detected in the field.



\section*{Acknowledgements}
This research has made use of the HASH PN database at hashpn.space. This research made use of APLpy, an open-source plotting package for Python \citep{Robitaille2012}.

\section*{Supporting information}
Additional Supporting Information may be found in the online version of this article:\\\\
\textbf{Online Appendix A}\\
\textbf{Online Appendix B}

\label{lastpage}

\end{document}